\documentclass[12pt]{article}
\usepackage{comment}
\usepackage{array}
\usepackage{booktabs} 
\usepackage{amssymb} 
\newcolumntype{C}[1]{>{\centering\arraybackslash}m{#1}}

\usepackage[T1]{fontenc}
\usepackage[utf8]{inputenc}
\usepackage{lmodern}

\usepackage{amsmath,amssymb,amsthm,mathtools,bm}
\usepackage{algorithm}
\usepackage{algpseudocode}
\usepackage{enumitem}
\usepackage{booktabs}
\usepackage{longtable}
\usepackage{placeins}
\usepackage{float}
\usepackage{array}
\usepackage{microtype}
\usepackage{xcolor}
\usepackage{xurl}
\usepackage[round,authoryear]{natbib}
\usepackage{xr-hyper}
\usepackage[colorlinks=true,allcolors=blue]{hyperref}
\usepackage[nameinlink,capitalize,noabbrev]{cleveref}
\hypersetup{allcolors=blue}
\newtheorem{theorem}{Theorem}[section]

\newtheorem{remark}[theorem]{Remark}

\crefname{assumption}{Assumption}{Assumptions}
\Crefname{assumption}{Assumption}{Assumptions}
\crefname{algorithm}{Algorithm}{Algorithms}
\Crefname{algorithm}{Algorithm}{Algorithms}

\def\E{\mathbb{E}}

\def\P{\mathbb{P}}
\def\o{{\scriptstyle{\mathcal{O}}}}

\newcommand{\Var}{\operatorname{Var}}

\newcommand{\argmin}{\operatorname*{arg\,min}}

\newcommand{\calX}{\mathcal X}

\newcommand{\Frechet}{Fr\'{e}chet}

\newcommand{\R}{\mathbb{R}}
\newcommand{\Om}{\Omega}

\newcommand{\Sym}{\mathrm{Sym}}

\newcommand{\Log}{\operatorname{Log}}
\newcommand{\Exp}{\operatorname{Exp}}

\begin{document}

\def\spacingset#1{\renewcommand{\baselinestretch}%
{#1}\small\normalsize} \spacingset{1}

%%%%%%%%%%%%%%%%%%%%%%%%%%%%%%%%%%%%%%%%%%%%%%%%%%%%%%%%%%%%%%%%%%%%%%%%%%%%%%
\date{}

  \title{\bf MATCH: Multiplier-Assisted Tests for Conditional Hypotheses in Non-Euclidean Data}
\author{Leheng Cai\footnotemark[1] \footnotemark[6], $\,$ Xu Guo,\footnotemark[4]   \footnotemark[5] \footnotemark[6]  $\,$ and$\,$  Qirui Hu\footnotemark[2] \footnotemark[3] \footnotemark[5]  \footnotemark[6]   \hspace{.2cm}}
  \renewcommand{\thefootnote}{\fnsymbol{footnote}}

  \footnotetext[1]{Department of Statistics and Data  Science, Tsinghua University}
     \footnotetext[2]{School of Statistics and Data Science, Shanghai University of Finance and Economics}
  \footnotetext[3]{Institute of Big Data Research, Shanghai University of Finance and Economics}
        \footnotetext[4]{School of Statistics, Beijing Normal University}
        \footnotetext[5]{Corresponding authors:  xustat12@bnu.edu.cn; huqirui@mail.shufe.edu.cn.}
        \footnotetext[6]{All authors   {contributed equally to} this article, and  are listed in   {alphabetical order}.}
\begin{comment}
          
\end{comment}

\bigskip
\maketitle
	\textbf{Abstract:} We propose a new procedure MATCH (\textbf{M}ultiplier-\textbf{A}ssisted \textbf{T}ests for \textbf{C}onditional \textbf{H}ypotheses) to test whether the non-Euclidean data match the target model, which is a general framework for significance and specification testing in \Frechet{} regression. MATCH covers global significance, partial significance, and the adequacy of global \Frechet{} regression, providing a unified way to compare unrestricted conditional \Frechet{} means with restricted alternatives. One of the key challenges is that the ordinary held-out loss difference is first-order degenerate under the null: the oracle losses coincide, and plug-in statistics are dominated by nuisance estimation error. MATCH uses sample splitting and independent random multipliers on
held-out losses to create a nondegenerate Gaussian leading term without residuals or tangent-space coordinates. To improve data use and stability, we further develop cross-fitted tests and repeated cross-fitting with \(p\)-value merging. We establish asymptotic null validity, consistency under fixed alternatives, and local power guarantees. Simulations for distributional, symmetric positive-definite (SPD) matrix-valued, and spherical responses support the theoretical findings, and applications to county-level household income distributions and North Atlantic tropical-cyclone locations demonstrate the practical use of the proposed tests.


\newpage
\spacingset{1.9}

\section{Introduction}\label{sec:introduction}

\begin{figure}[h!]
    \centering
    \includegraphics[width=0.45\linewidth]{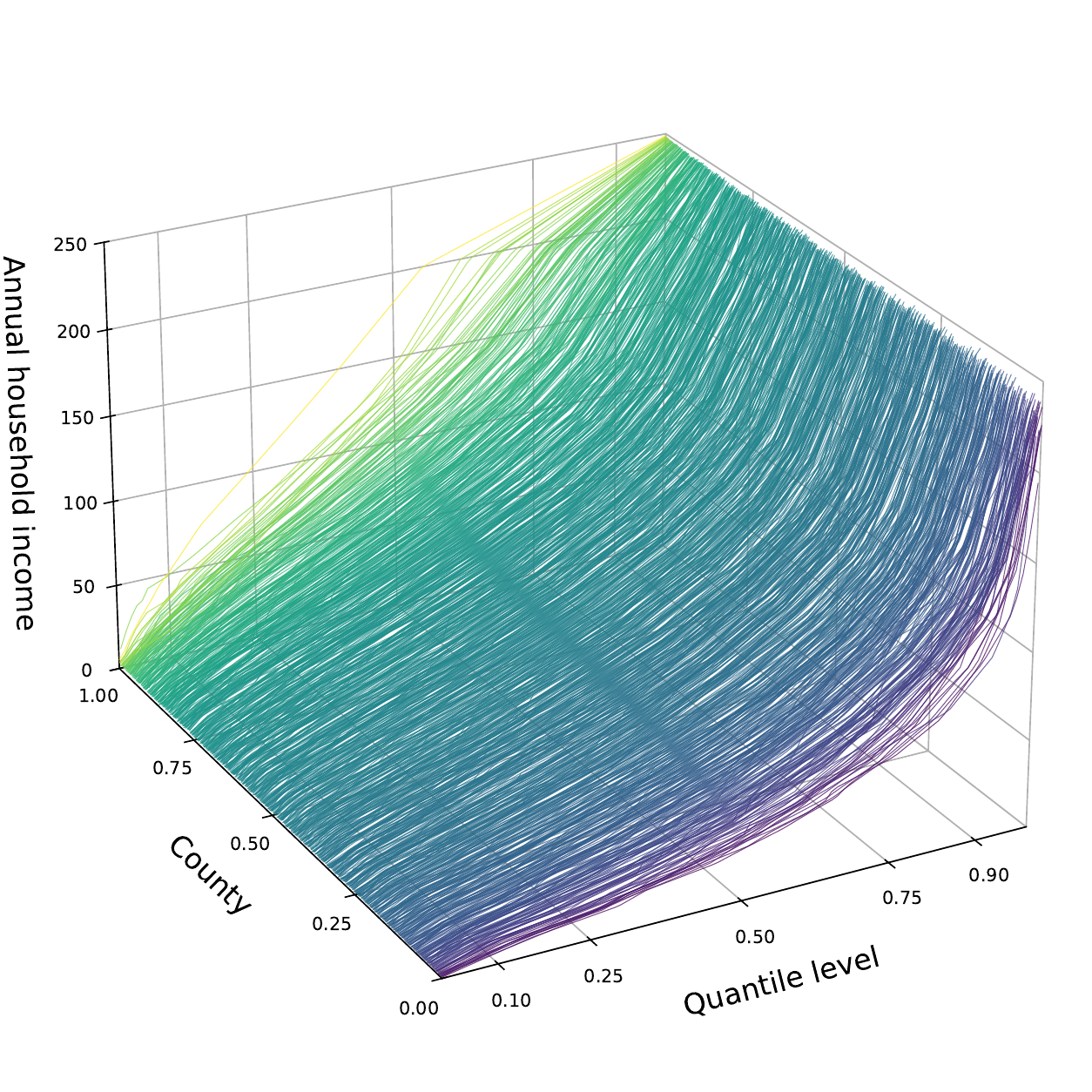}
    \includegraphics[width=0.4\linewidth]{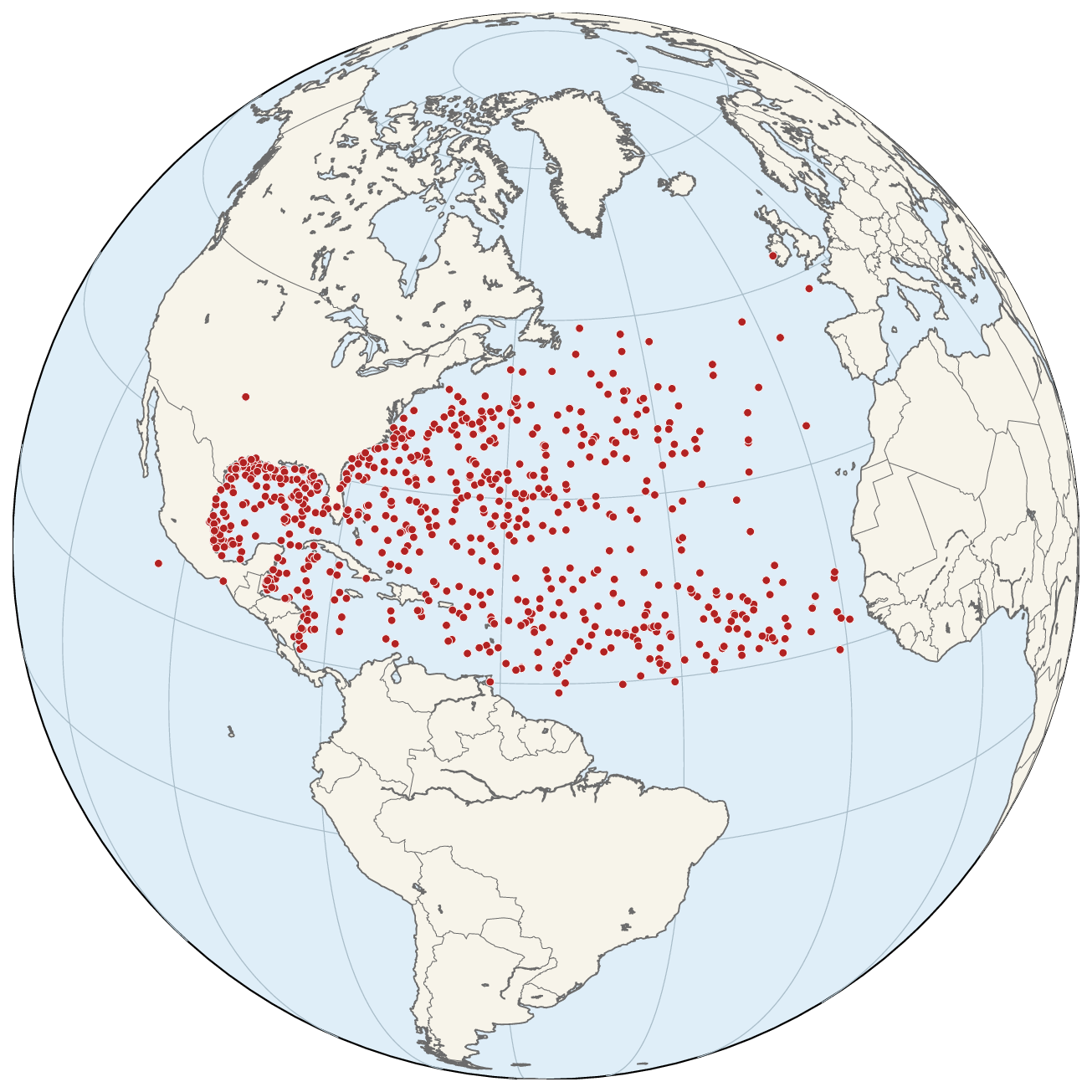}
    \caption{Left panel: smoothed household-income quantile functions for 1000 randomly selected U.S. counties from 2020--2024. Right panel: locations of lifetime maximum intensity for North Atlantic tropical cyclones from 1980--2025.}
    \label{fig:example}
\end{figure}

Modern statistical data analysis increasingly involves non-Euclidean data objects whose intrinsic structure is not well represented by ordinary Euclidean vectors. 
Examples include probability distributions endowed with optimal-transport metrics, covariance and correlation matrices constrained to positive-definite cones, networks represented by graph objects or Laplacians, shapes modulo translation and rotation, tree-structured data, compositional vectors on simplices, and observations on spheres. 
Figure~\ref{fig:example} provides motivating examples of non-Euclidean data objects, including household-income distributions for U.S. counties and the locations at which North Atlantic tropical cyclones attain their lifetime maximum intensity; details are given in Section~\ref{sec:realdata}.

The object-oriented data analysis perspective emphasizes that such observations should often be treated as data objects with their own geometry, rather than as unconstrained vectors after ad hoc vectorization \citep{MarronAlonso2014,MarronDryden2021}. Related developments include statistics on manifolds \citep{BhattacharyaPatrangenaru2003,BhattacharyaPatrangenaru2005,PatrangenaruEllingson2015}, statistical shape analysis \citep{DrydenMardia2016}, and statistics in Wasserstein space \citep{PanaretosZemel2020}. A common feature of these settings is that addition, scalar multiplication, residual vectors, and linear projections may be unavailable or geometrically inappropriate.

\Frechet{} mean, introduced by \citet{Frechet1948}, provides a basic analogue of the Euclidean expectation in a metric space. If \(Y\) is a random object in \((\Omega,d)\), its \Frechet{} mean is
\[
\omega_0:=\argmin_{\omega\in\Omega}\E\{d^2(Y,\omega)\}.
\]
\citet{PetersenMuller2019} extended this idea to regression with Euclidean predictors \(\bm X\in\calX\subseteq\mathbb R^p\) by defining the conditional \Frechet{} objective
\[
M(\omega,\bm x)=\E\{d^2(Y,\omega)\mid \bm X=\bm x\},\qquad \omega\in\Omega,
\]
and the conditional \Frechet{} mean \(m(\bm x)=\argmin_{\omega\in\Omega}M(\omega,\bm x)\). They introduced global \Frechet{} regression by rewriting ordinary least squares as a weighted \Frechet{} mean, and local \Frechet{} regression by replacing local smoothing averages with local weighted \Frechet{} means. Subsequent work has developed uniform convergence theory for local \Frechet{} regression \citep{ChenMuller2022}, nonparametric regression in nonstandard spaces including local-linear and series constructions \citep{Schoetz2022}, random-forest-weighted local \Frechet{} regression \citep{QiuYuZhu2024}, \Frechet{} random forests for metric-space-valued regression with non-Euclidean predictors \citep{CapitaineBigotThiebautGenuer2024}, dimension reduction and variable selection for \Frechet{} regression \citep{ying2022frechet, ZhangXueLi2024,TuckerWuMuller2023}, and deep \Frechet{} regression for high-dimensional Euclidean predictors \citep{IaoZhouMuller2025}. These developments make the conditional \Frechet{} mean a central regression target for complex random objects.

In Euclidean regression, specification and significance testing for conditional means has a long history. Classical questions include whether a parametric or linear model adequately represents a nonparametric regression function, and whether a subset of covariates has no additional mean effect. Representative contributions include tests comparing parametric and nonparametric fits \citep{HardleMammen1993}, kernel-based functional-form tests \citep{Zheng1996}, reduction-based adaptive-to-model tests \citep{tan2019adaptive,tan2022integrated}, omitted-variable tests \citep{FanLi1996}, bootstrap significance tests in nonparametric regression \citep{DelgadoGonzalezManteiga2001}, and recent machine-learning and sample-splitting tests for partial mean dependence \citep{lundborg2022projected,williamson2023general,dai2022significance,CaiGuoZhong2025,zhang2025testing}. These procedures typically rely on scalar residuals, inner products, or linear projections, which are not available for a generic metric-space response.

Inference for \Frechet{} regression remains less developed than estimation. In special response geometries, \citet{PetersenLiuDivani2021} developed Wasserstein \(F\)-tests and confidence bands for density response curves, while \citet{XuLi2025Partial, XuLi2025Bures} studied inference for covariance responses under Bures-Wasserstein geometry. Recent work has also begun to address inference for \Frechet{} regression in broader metric spaces \citep{SongDubeyMullerPetersen2026}. These contributions provide important benchmarks, but existing inference procedures are either tied to particular response geometries or to more specific \Frechet{} regression targets.

In this paper, we develop significance and specification tests for \Frechet{} regression within a general restricted-versus-unrestricted framework by MATCH procedures. 
Specifically, for a given objective function \(G(\omega,\bm x)\), we aim to test
\begin{equation}\label{EQ:H0}
    H_0:\mathbb P\left(m(\bm X)=g(\bm X)\right)=1,\qquad
    H_1:\mathbb P\left(m(\bm X)\neq g(\bm X)\right)>0,
\end{equation}
where \(g(\bm x):=\argmin_{\omega\in\Omega}G(\omega,\bm x)\).

This conditional-hypothesis formulation encompasses several important testing problems.

\begin{itemize}
    \item[(i)] \textbf{Testing global significance.}
    When \(G(\omega,\bm x)\equiv\E\{d^2(Y,\omega)\}\), the function \(g(\bm x)\equiv\omega_0\) is the unconditional \Frechet{} mean. The resulting hypothesis tests whether the conditional \Frechet{} mean is almost surely equal to the unconditional \Frechet{} mean:
    \begin{equation}\label{eq:H0_global}
    H_0:\mathbb P\left(m(\bm X)=\omega_0\right)=1,\qquad
    H_1:\mathbb P\left(m(\bm X)\neq\omega_0\right)>0.
    \end{equation}

    \item[(ii)] \textbf{Testing partial significance.}
    Write \(\bm X=(\bm Z^\top,\bm W^\top)^\top\) and \(\bm x=(\bm z^\top,\bm w^\top)^\top\). When \(G(\omega,\bm x)=\E\{d^2(Y,\omega)\mid \bm Z=\bm z\}\), \(g(\bm x)=f(\bm z)\) is the partial \Frechet{} mean conditional on \(\bm Z\). The test asks whether \(\bm W\) has any additional effect on the conditional \Frechet{} mean beyond that explained by \(\bm Z\). This target is weaker than conditional independence, because the conditional distribution of \(Y\) may depend on \(\bm W\) through dispersion or other higher-order features while leaving the conditional \Frechet{} mean unchanged.

    \item[(iii)] \textbf{Testing the global \Frechet{} regression specification.}
    Let \(\bm \mu=\E\bm X\), let \(\mathbf\Sigma=\Var(\bm X)\) be nonsingular, and define
    \(s(\bm x',\bm x)=1+(\bm x'-\bm\mu)^\top\mathbf\Sigma^{-1}(\bm x-\bm\mu)\).
    When \(G(\omega,\bm x)=\E\{s(\bm X,\bm x)d^2(Y,\omega)\}\), the restricted target is the global \Frechet{} regression function \(m_0(\bm x)\).  The corresponding conditional hypothesis asks whether the unrestricted conditional \Frechet{} mean equals this global \Frechet{} regression restricted target. In Euclidean squared-loss regression, this reduces to testing a linear conditional-mean specification.
\end{itemize}

The main difficulty in testing the conditional hypothesis \eqref{EQ:H0} is that ordinary residual-based testing has no direct metric-space analogue. 
Moreover, under \(H_0\), the oracle held-out loss difference $
    d^2(Y,m(\bm X))-d^2(Y,g(\bm X))$
is identically zero. After \(m(\cdot)\) and \(g(\cdot)\) are replaced by estimators, a naive held-out loss comparison is therefore driven by nuisance-estimation error rather than by a nondegenerate empirical-process term. We address this degeneracy by using sample splitting and independent random multipliers. The training sample estimates the unrestricted and restricted \Frechet{} regression functions, while the validation sample is used only to compute randomized held-out losses. Under the null, the oracle losses coincide, but the difference between independent multiplier perturbations creates a nondegenerate leading term. The resulting statistic can be calibrated by a Gaussian approximation and further improved by cross-fitting.

The paper makes three main contributions. First, it frames global significance, partial significance, and global \Frechet{} specification testing as conditional hypotheses comparing an unrestricted conditional \Frechet{} mean with a restricted target. Second, it introduces the MATCH construction: multiplier-assisted held-out loss statistics that avoid residual vectors and tangent-space coordinates while producing a nondegenerate null limit. Third, it develops MATCH procedures with cross-fitted and repeated cross-fitted refinements, establishes null validity and consistency against fixed alternatives, and rigorously proves that the cross-fitted procedure attains strictly higher local asymptotic power than its single-split counterpart.
Table~\ref{tab:related_work_comparison} compares related work with representative references listed in the table notes. Existing Euclidean tests address related conditional-mean questions in vector spaces, while available \Frechet{}-regression inference is either tied to special response geometries or to more specific inferential targets. To our knowledge, this is the first general testing framework that directly compares unrestricted and restricted conditional \Frechet{} means in metric spaces.

%The present paper develops a unified restricted-versus-unrestricted testing framework for conditional \Frechet{} means, so that common significance tests, such as, global significance, partial significance, and global \Frechet{} specification testing, are treated as special instances of the same comparison.

\begin{table}[!htbp]
\centering
\caption{Comparison with related works.}
\label{tab:related_work_comparison}
\scriptsize
\setlength{\tabcolsep}{2.6pt}
\renewcommand{\arraystretch}{0.88}
\begin{tabular}{@{} p{0.41\textwidth}cccc@{}}
\toprule
Representative works
& \begin{tabular}{@{}c@{}}General metric-\\space  response\end{tabular}
& Inference
&  Model-free 
& \begin{tabular}{@{}c@{}}Unified\\sig./spec.\end{tabular}
\tabularnewline
\midrule
Specification tests for Euclidean responses$^{1}$
& \(\times\) & \(\checkmark\) & \(\checkmark\) & \(\times\) \\
Significance tests for Euclidean responses$^{2}$ & \(\times\) & \(\checkmark\) & \(\checkmark\) & \(\times\)\\
Unified tests for Euclidean responses$^{3}$ & \(\times\) & \(\checkmark\) & \(\checkmark\) & \(\checkmark\) \\
Conditional \Frechet{} mean estimation$^{4}$  & \(\checkmark\) & \(\times\) & \(\times\) & \(\times\) \\
Variable selection/dimension reduction$^{5}$
& \(\checkmark\) & \(\times\) & \(\times\) & \(\times\) \\
 Significance tests for distributional/SPD responses$^{6}$ & \(\times\) & \(\checkmark\) & \(\times\) & \(\times\) \\
 Significance tests for metric-space responses$^{7}$
& \(\checkmark\) & \(\checkmark\) & \(\times\) & \(\times\) \\
 Unified tests for metric-space responses$^{8}$
& \(\checkmark\) & \(\checkmark\) & \(\checkmark\) & \(\checkmark\) \\
\bottomrule
\end{tabular}
\vspace{0.25em}
\begin{minipage}{0.98\textwidth}
\tiny
\emph{Notes:} \(^{1}\)\citet{HardleMammen1993,Zheng1996,tan2019adaptive,tan2022integrated};
\(^{2}\)\citet{DelgadoGonzalezManteiga2001,lundborg2022projected, williamson2023general,dai2022significance,CaiGuoZhong2025,zhang2025testing};
\(^{3}\)\citet{FanLi1996};
\(^{4}\)\citet{PetersenMuller2019,ChenMuller2022,Schoetz2022, QiuYuZhu2024,IaoZhouMuller2025};
\(^{5}\)\citet{TuckerWuMuller2023,ZhangXueLi2024}; \(^{6}\)\citet{PetersenLiuDivani2021,XuLi2025Bures,XuLi2025Partial};\(^{7}\)\citet{SongDubeyMullerPetersen2026}; \(^{8}\) This paper.
\end{minipage}
\end{table}

The remainder of the paper is organized as follows. Section~\ref{sec:method} introduces the MATCH procedure and establishes the corresponding asymptotical properties. Section~\ref{sec:crossfitting} develops the cross-fitted and repeated cross-fitted versions and gives the local power comparison. Sections~\ref{sec:simulation} and \ref{sec:realdata} present simulation studies and real-data applications, respectively. Additional simulations, discussion of technical assumptions, and detailed proofs are provided in the Supplementary Material.

\section{Main results}\label{sec:method}

Following \cite{PetersenMuller2019}, we assume that the metric space $(\Omega,d)$ is separable and has  finite diameter $D<\infty$.

\subsection{Degeneracy issue under the null}

A key difficulty in testing
\eqref{EQ:H0}
is that the naive loss-difference statistic
\begin{align*}
     S_n= \frac1n\sum_{i=1}^n
    \left\{
        d^2(Y_i,\widehat m(\bm X_i))
        -
        d^2(Y_i,\widehat g(\bm X_i))
    \right\},
\end{align*}
is degenerate under the null, where $\widehat m(\cdot)$ and $\widehat g(\cdot)$ are some consistent estimators of $m(\cdot) $ and $ g(\cdot)$.
To explain this, consider the oracle test statistic
\[
   \widetilde S_n= \frac1n\sum_{i=1}^n
    \left\{
        d^2(Y_i,m(\bm X_i))
        -
        d^2(Y_i,g(\bm X_i))
    \right\}.
\]
Under \(H_0\), \(m(\cdot)=g(\cdot)\) almost surely, and thus each summand is exactly
zero, implying $\widetilde S_n\equiv 0$ almost surely. Therefore, the usual root-\(n\) fluctuation of \(\widetilde S_n\) disappears.
If \(m\) and \(g\) are replaced by their estimators accordingly, the resulting naive statistic \(  S_n\) is then
driven mainly by nuisance estimation errors rather than by an ordinary empirical
process term. This leads to a degenerate or nonstandard limiting behavior and
would require a much more delicate analysis of the estimation errors.
Such an analysis is already highly nontrivial even for
global linear Fréchet regression in general non-Euclidean metric spaces \citep{PetersenMuller2019}, and the difficulty becomes even more pronounced for other nonparametric   Fréchet
regression \citep{IaoZhouMuller2025}. %, and for more flexible estimators such as deep Fréchet regression \citep{IaoZhouMuller2025}, whose estimation errors are typically hard  to characterize.

\subsection{MATCH procedure}

The MATCH procedure addresses this difficulty by applying independent multipliers to held-out losses, thereby creating a nondegenerate asymptotic distribution under the null while requiring only mild conditions on the nuisance estimation errors.

We first split the data \(\mathcal D=\{\bm X_i, Y_i\}_{i=1}^n\) into two disjoint subsets
\(\mathcal D_1\) and \(\mathcal D_2\), with $n_1=|\mathcal D_1|$, $n_2=|\mathcal D_2|$. On \(\mathcal D_1\), we construct the estimators
$\widehat m_{\mathcal D_1}(\cdot)$  and $\widehat g_{\mathcal D_1}(\cdot)$.
The held-out sample \(\mathcal D_2\) is   used for inference. %On $\mathcal D_1$, train $\widehat m_{\mathcal D_1}(\cdot)$  and $\widehat g_{\mathcal D_1}(\cdot)$.
%Then, we use $\mathcal D_2$ to make inference.
Then, we generate independent multiplier pairs $(\varphi_{i},\psi_{i})_{i\in \mathcal D_2}$, independent of the data $\mathcal D$, satisfying
\begin{equation*}\label{eq:multiplier_moments}
\E\varphi_{i}=\E\psi_{i}=1,
\quad
\Var(\varphi_{i})=\Var(\psi_{i})=1/2,\quad \E|\varphi_{i}|^{2+\delta}+ \E|\psi_{i}|^{2+\delta}<\infty,\quad\delta>0.
\end{equation*}
Let $\widetilde T_n=T_n-\max\{b_n,0\}$, where \begin{align*}
    &T_{n}=\frac{1} {n_2}\sum_{i\in \mathcal D_2}\left\{\varphi_{i }d^2(Y_i,\widehat m_{\mathcal D_1}(\bm X_i)  )-\psi_{i }d^2(Y_i,\widehat g_{\mathcal D_1}(\bm X_i)  )\right\},\\&b_n = \frac{1} {n_2}\sum_{i\in \mathcal D_2}\left\{ d^2(Y_i,\widehat m_{\mathcal D_1}(\bm X_i)  )- d^2(Y_i,\widehat g_{\mathcal D_1}(\bm X_i)  )\right\}.
\end{align*}
Here, $\max\{b_n, 0\}$ is a finite-sample correction term, which will be further discussed in Remark \ref{bn}. 
Under $H_0$, the null variance is estimated by
\begin{equation}\label{eq:sigmahat}
\widehat\sigma_n^2=\frac1{ n_2}\sum_{i\in\mathcal D_2} d^4(Y_i,\widehat m_{\mathcal D_1}(\bm X_i)  ) .
\end{equation}

We make some intuitive explanations for our MATCH procedure. Under assumptions stated later, we can show that 
\begin{align*}
    &T_{n}=\frac{1} {n_2}\sum_{i\in \mathcal D_2}\left\{\varphi_{i }d^2(Y_i, m(\bm X_i)  )-\psi_{i }d^2(Y_i, g(\bm X_i)  )\right\}+\o_p\left(\frac{1}{\sqrt{n_2}}\right).
\end{align*}
Under \(H_0\), \(m(\cdot)=g(\cdot)\) almost surely,   thus 
\begin{align*}
    &T_{n}=\frac{1} {n_2}\sum_{i\in \mathcal D_2}\left\{(\varphi_{i }-\psi_{i })d^2(Y_i, g(\bm X_i)  )\right\}+\o_p\left(\frac{1}{\sqrt{n_2}}\right).
\end{align*}
Since the multiplier pairs $(\varphi_{i})_{i\in \mathcal D_2}$ and $(\psi_{i})_{i\in \mathcal D_2}$ are independent and also independent of the data $\mathcal D$, the random variables  $(\varphi_{i }-\psi_{i })d^2(Y_i, g(\bm X_i))$'s have zero expectation and finite variance. This then leads $T_n$ to avoid the degeneracy issue under the null. On the other hand, the  leading term of $T_n$ has expectation $\E\left\{d^2(Y_i, m(\bm X_i)  )-d^2(Y_i, g(\bm X_i)  )\right\}$ under $H_1$,  which is strictly negative. This implies that we can use $T_n$ and its modified version $\widetilde T_n$ to test the hypothesis \eqref{EQ:H0}. 

We state the assumptions used to derive the asymptotic behavior of the studentized statistic \(\sqrt{n_2}\widetilde T_n/\widehat\sigma_n\).
\begin{itemize}
\item[(A0)] For some $\rho\in(0,1)$, $\lim_{n\to\infty}n_1/n= \rho$.
    \item[(A1)] The functions $\omega\mapsto M(\omega,\bm x)$ and $\omega\mapsto G(\omega,\bm x)$  are uniquely minimized at  $m(\bm x)$ and $g(\bm x)$,  respectively. There exist positive constants $\delta_0$ and $C$ such that, for any $\bm x\in\mathcal X$,  \begin{align*}
        &0\leq M(\omega,\bm x)-M(m(\bm x),\bm x)\leq C d^2(\omega,m(\bm x)),\quad \forall d(\omega,m(\bm x)) \leq \delta_0\\
        &0\leq G(\omega,\bm x)-G(g(\bm x),\bm x)\leq C d^2(\omega,g(\bm x)),\quad \forall d(\omega,g(\bm x)) \leq \delta_0.
    \end{align*}
    \item[(A2)] The estimation errors  {satisfy} \begin{align*}
        &\E \left\{ d ^2 ( \widehat m_{\mathcal D_1}(\bm X),m(\bm X)) |\mathcal D_1 \right\} =\o_p\left(n^{-1/2}\right),\,\,\, \E \left\{ d  ^2 ( \widehat g_{\mathcal D_1}(\bm X),g(\bm X)) |\mathcal D_1 \right\} =\o_p\left(n^{-1/2}\right),
    \end{align*}
    where \((\bm X,Y)\) is an independent copy of the observation.
\end{itemize}

Assumption (A1) serves as a local smoothness condition for the Fréchet
objectives. It holds automatically in Euclidean spaces, where the excess
squared-loss risk admits an exact quadratic expansion around the conditional
mean.  For general metric spaces, however, where no linear structure is available, this
condition provides a metric-space analogue of the local second-order Taylor
expansion used in the Euclidean setting. In Section~\ref{sec:A1}, we verify and discuss
  this
assumption for several representative non-Euclidean spaces.

The condition in Assumption (A2) for the estimator \(\widehat m_{\mathcal D_1}(\cdot)\) of the unrestricted target
{can be verified for standard local \Frechet{} regression estimators under appropriate dimension, smoothness, and bandwidth conditions; see \citet{PetersenMuller2019}.}
This is also the default estimator
used for \(m(\cdot)\) in our simulations. The requirement on $\widehat g_{\mathcal D_1}(\cdot)$   is a high-level condition, which depends on
 the form of the restricted objective \(G(\omega,\bm x)\). In the global
significance test, \(g(\cdot)\) is estimated by the sample \Frechet{} mean; in the
partial significance test, \(g(\cdot)\) can be estimated by local \Frechet{} regression
based on the reduced covariates; and in the global \Frechet{} specification test, \(g(\cdot)\) is estimated
by global linear \Frechet{} regression.
Under  standard regularity conditions
for these estimators, the requirement in Assumption (A2) is fulfilled; see \cite{BhattacharyaPatrangenaru2003,BhattacharyaPatrangenaru2005,PetersenMuller2019}.
Note that these conditions coincide with those imposed in
\cite{10.1111/ectj.12097,williamson2023general} in the Euclidean setting.

We emphasize that sample splitting is essential for the general  validity of the proposed method under the stated high-level   conditions in Assumption (A2). Theoretically,  
establishing validity without splitting would  require additional learner-specific assumptions, such as empirical-process restrictions in  \cite{SongDubeyMullerPetersen2026}.
From a practical perspective, if the same observations were used for both fitting and testing, the more flexible unrestricted learner may achieve a smaller in-sample loss than the restricted learner even under the null, inducing a negative loss gap in the rejection direction and causing size distortion.

The  {nondegeneracy} of the proposed single-split multiplier statistic \(\widetilde T_n\)
is theoretically guaranteed by Theorem~\ref{THM:H0}.
\begin{theorem}\label{THM:H0}
 Under $H_0$, suppose that Assumptions (A0)-(A2) hold and $\E d^4(Y,m(\bm X))>0$. Then, as $n\to\infty$, \begin{equation*}
 \sqrt{n_2} \widetilde T_{n}/\widehat\sigma_n\xrightarrow{d}\mathcal N(0,1).
\end{equation*}
\end{theorem}

 Theorem \ref{THM:H0} shows that  the test that rejects \(H_0\) when
\(
{\sqrt{n_2}\widetilde T_n}/{\widehat\sigma_n}<z_\alpha
\)
has asymptotic level \(\alpha\), where \(z_\alpha\) is the \(\alpha\)-quantile
of the standard Gaussian distribution. Equivalently, the one-sided \(p\)-value of the proposed test  is given by
\(\Phi({\sqrt{n_2}\widetilde T_n}/{\widehat\sigma_n}),
\)
  where $\Phi(\cdot)$ is the cumulative distribution function of the standard Gaussian random variable.

\begin{remark}\label{bn}
     The conclusion of Theorem~\ref{THM:H0} remains unchanged if
\(\widetilde T_n\) is replaced by the original statistic \(T_n\),  since $\sqrt{n_2}b_n$ is asymptotically negligible under the null. %Therefore,  \(T_n\) is also asymptotically valid.
However, we find that the empirical sizes of the test based directly
on \(T_n\) are often conservative. This phenomenon  {mainly occurs because the}
unrestricted estimator \(\widehat m_{\mathcal D_1}(\cdot)\) may have a larger
held-out loss than \(\widehat g_{\mathcal D_1}(\cdot)\) in finite samples,  though \(m(\cdot  )=g(\cdot)\) at the population level under the null.
% the
% finite-sample plug-in bias induced by estimating the unrestricted conditional
% \Frechet{} mean with a more flexible local estimator than the restricted target
% under the null. Although \(m(\cdot  )=g(\cdot)\) at the population level, the
% unrestricted estimator \(\widehat m_{\mathcal D_1}(\cdot)\) may have a larger
% held-out loss than \(\widehat g_{\mathcal D_1}(\cdot)\) in finite samples.
As a
result, \(b_n\) tends to be slightly positive under the null, which shifts
the left-sided statistic to the right and leads to conservative rejection
probabilities.
 Subtracting \(\max\{ b_n,0\}\)
%  , which  can be viewed as a finite-sample estimate of the
% positive part of this conditional plug-in drift,
   provides a simple bias correction, reducing the rightward shift of
the null distribution caused by nuisance estimation errors. The left panel of
Figure~\ref{fig:hist_diagnostics} illustrates this finite-sample correction in
the distributional simulation.

\end{remark}

\begin{remark}
    Another natural variance estimator is
\begin{equation} \label{DEF:sigmatilde}
\begin{aligned}
\widetilde\sigma_n^2&=\frac1{2n_2}\sum_{i\in\mathcal D_2}\left\{d^4(Y_i,\widehat m_{\mathcal D_1}(\bm X_i)  )+d^4(Y_i,\widehat g_{\mathcal D_1}(\bm X_i)  )\right\}\\&+
\frac1{n_2}\sum_{i\in\mathcal D_2}\left\{d^2(Y_i,\widehat m_{\mathcal D_1}(\bm X_i)  )-d^2(Y_i,\widehat g_{\mathcal D_1}(\bm X_i)  )-b_n\right\}^2.
\end{aligned}
\end{equation}
This estimator is consistent for the asymptotic variance of $\sqrt{n_2}\widetilde T_n$, i.e.,  
$$\left\{ \E d^4(Y,m(\bm X))
+
  \E d^4(Y,g(\bm X))\right\}/2+\Var\left( d^2(Y,m(\bm X))-d^2(Y,g(\bm X)) \right),$$
 under both the null and   alternatives.
Under alternatives, however, 
\[
\widetilde\sigma_n^2-\widehat\sigma_n^2
\xrightarrow{\P}
\frac12
\left\{
\E d^4(Y,g(\bm X))
-
\E d^4(Y,m(\bm X))
\right\}+ \Var\left( d^2(Y,m(\bm X))-d^2(Y,g(\bm X)) \right).
\]
Since \(g(\cdot)\) is a restricted target under the null model,   one often expects
$\E d^4(Y,g(\bm X))
\geq
\E d^4(Y,m(\bm X))$,
in which case \(\widetilde\sigma_n^2\) has a larger probability limit than
\(\widehat\sigma_n^2\). Consequently, studentization by
\(\widetilde\sigma_n\) may be more conservative under alternatives, whereas
\(\widehat\sigma_n\)   yields higher power while preserving the same asymptotic distribution under the null. The right panel of
Figure~\ref{fig:hist_diagnostics} compares the  studentized statistics $\sqrt{n_2}\widetilde T_n/\widehat\sigma_n$ and $\sqrt{n_2}\widetilde T_n/\widetilde\sigma_n$ 
under an alternative.
\begin{figure}[h!]
    \centering
    \begin{minipage}{0.49\linewidth}
        \centering
        \includegraphics[width=\linewidth]{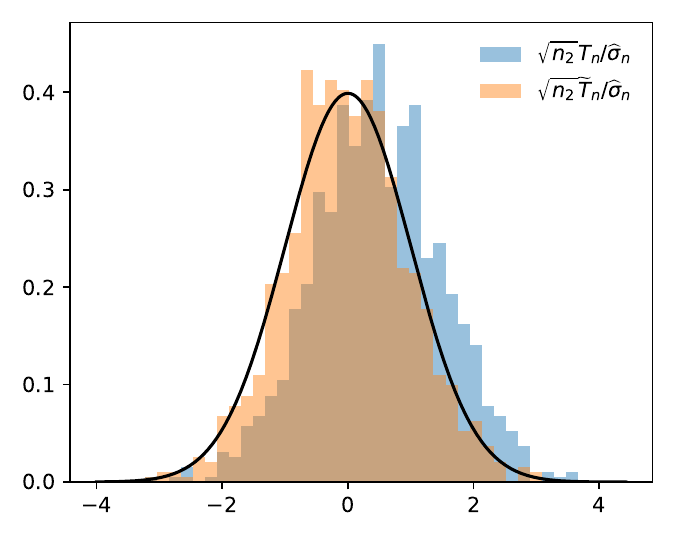}
        {\small (a) Finite-sample correction.}
    \end{minipage}
    \begin{minipage}{0.49\linewidth}
        \centering
        \includegraphics[width=\linewidth]{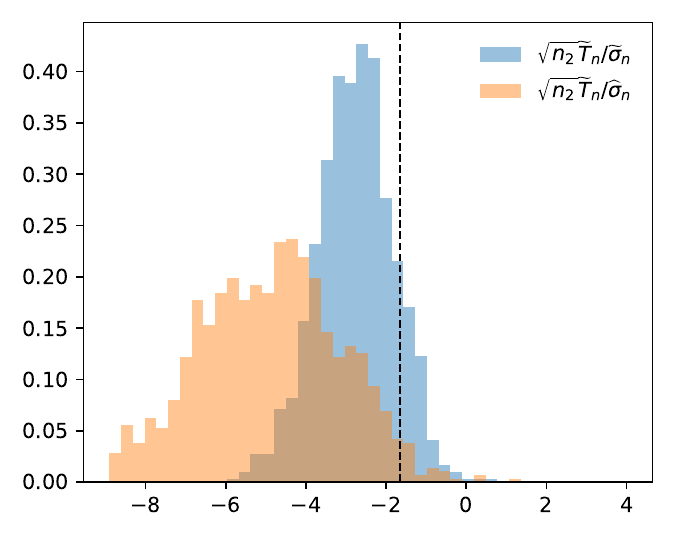}
        {\small (b) Two studentized statistics under $H_1$.}
    \end{minipage}
    \caption{Diagnostic histograms for the single-split statistic in the distributional-response simulation of Section~\ref{sec:simulation-dist}. Left: statistics $\sqrt{n_2}T_n/\widehat\sigma_n$ and $\sqrt{n_2}\widetilde T_n/\widehat\sigma_n$ under the global significance null hypothesis \eqref{eq:H0_global} with $n=400$; the black solid curve is the standard normal density. Right: statistics $\sqrt{n_2}\widetilde T_n/\widehat\sigma_n$ and $\sqrt{n_2}\widetilde T_n/\widetilde\sigma_n$ under the corresponding alternative with $n=400$ and $a=0.05$; the black solid vertical line is the critical value \(z_{\alpha}\) at level \(\alpha=0.05\).}
    \label{fig:hist_diagnostics}
\end{figure}

\end{remark}

\begin{remark}\label{remark2}
    Another asymptotically valid statistic  is
\begin{equation*}%\label{eq:Tnk}
T_{n}^{\dagger}=\frac{1} {n_2}\sum_{i\in \mathcal D_2}\left\{\varphi'_{i }d^2(Y_i,\widehat m_{\mathcal D_1}(\bm X_i)  )- d^2(Y_i,\widehat g_{\mathcal D_1}(\bm X_i)  )\right\}-\max\{b_n,0\},
\end{equation*}
where $\E\varphi'_{i} =1$,
and
$\Var(\varphi'_{i}) =1 $, and the same variance estimator \(\widehat\sigma_n^2\) is applicable.
This asymmetric version avoids randomizing the restricted loss term, and its finite-sample
performance is similar to that of the  symmetric multiplier statistic $\widetilde T_n$; see additional numerical studies in Section \ref{sec:additional-sim}.  %Their finite-sample performance is similar.
\end{remark}

For fixed alternatives, we replace Assumption (A2) by the following much weaker consistency
Assumption (A2').

\begin{itemize}
   \item[(A2')] the estimation errors  {satisfy} \begin{align*}
        &\E \left\{ d^2  ( \widehat m_{\mathcal D_1}(\bm X),m(\bm X)) |\mathcal D_1 \right\} =\o_p\left(1\right),\,\,\,\E \left\{ d^2  ( \widehat g_{\mathcal D_1}(\bm X),g(\bm X)) |\mathcal D_1 \right\} =\o_p\left(1\right),
    \end{align*}
    where \((\bm X,Y)\) is an independent copy of the observation.
\end{itemize}
 Theorem \ref{THM:H1}  establishes the consistency of the proposed test {under fixed alternatives}.
\begin{theorem}\label{THM:H1}
Let $\Delta
=
\E\left\{
d^2(Y,m(\bm X))-d^2(Y,g(\bm X))
\right\}.$
Suppose that Assumptions (A0) and  (A2') hold, $ \E d^4(Y,m(\bm X))>0$,  and   {$ \Delta<0$},    as $n\to\infty$,
\[
\sqrt{n_2}\widetilde T_n/\widehat\sigma_n\xrightarrow{\P}-\infty.
\]
\end{theorem}

\section{Power enhancement and stabilization  procedures}\label{sec:crossfitting}
\subsection{Cross-fitted MATCH procedure}
%\subsection{$K$-fold cross-fitting tests}
%\textbf{Power enhancement}

To improve data efficiency, MATCH can be implemented through \(K\)-fold cross-fitting \citep{fan2012variance,10.1111/ectj.12097,vansteelandt2022assumption}, so that each observation is evaluated by nuisance estimators trained on the complementary folds.

For fixed $K\geq 2$, let \(\{\mathcal I_k\}_{k=1}^K\) be a partition of
\(\{1,\ldots,n\}\) into \(K\) disjoint folds with
\(|\mathcal I_k|=n_k\). For each \(k=1,\ldots,K\), define
\(
    \mathcal D_{-k}
    =
    \{(\bm X_i^\top ,Y_i)^\top: i\notin \mathcal I_k\}.
\)
Using \(\mathcal D_{-k}\), one constructs the nuisance estimators
$\widehat m_{-k}(\cdot)$ and $\widehat g_{-k}(\cdot)$.
Thus, each observation \(i\in\mathcal I_k\) is evaluated only by estimators
trained without using this observation.
% Generate independent multiplier pairs
% \((\varphi_i,\psi_i)_{i=1}^n\), independent of the data, satisfying
% \[
%     E\varphi_i=E\psi_i=1,\quad
%     \Var(\varphi_i)=\Var(\psi_i)=1/2,\quad E|\varphi_i|^{2+\delta}+E|\psi_i|^{2+\delta}<\infty,
% \]
% for some \(\delta>0\).
Generate independent multiplier pairs $(\varphi_i,\psi_i)_{i=1}^n$, independent of the data and folds, satisfying the same moment conditions as in the single-split procedure.
Then, the cross-fitting test statistic is defined as $\widetilde T_{n,K}= T_{n,K}-\max\{b_{n,K},0\}$,
where \begin{align*}
    &  T_{n,K}
    =
    \frac1n
    \sum_{k=1}^K
    \sum_{i\in\mathcal I_k}
    \left\{
        \varphi_i d^2(Y_i,\widehat m_{-k}(\bm X_i))
        -
        \psi_i d^2(Y_i,\widehat g_{-k}(\bm X_i))
    \right\},\\
    &  b_{n,K}
    =
    \frac1n
    \sum_{k=1}^K
    \sum_{i\in\mathcal I_k}
    \left\{
       d^2(Y_i,\widehat m_{-k}(\bm X_i))
        -
       d^2(Y_i,\widehat g_{-k}(\bm X_i))
    \right\}.
\end{align*}
Under the null hypothesis, the variance is estimated by
\[
    \widehat\sigma_{n,K}^2
    =
    \frac1n
    \sum_{k=1}^K
    \sum_{i\in\mathcal I_k}
    d^4(Y_i,\widehat m_{-k}(\bm X_i)).
\]
At significance level \(\alpha\), one rejects \(H_0\) if
\(  \sqrt{n}\widetilde T_{n,K}/\widehat\sigma_{n,K}<z_\alpha,
\)
where \(z_\alpha\) is the lower \(\alpha\)-quantile of the standard Gaussian
distribution.

We next state the asymptotic validity of the cross-fitted MATCH procedure.

% We assume that \(K\) is fixed and that the fold sizes are asymptotically balanced.

\begin{itemize}
\item[(B0)]    For each \(k=1,\ldots,K\), $\lim_{n\to\infty}n_k/n=\pi_k$, where  $0<\pi_k<1$ and $\sum_{k=1}^K\pi_k=1$.
\item[(B1)] For \(k=1,\ldots,K\), the cross-fitted  estimators  {satisfy}
\[
    \E\left\{
        d^2\left(\widehat m_{-k}(\bm X),m(\bm X)\right)
        \mid \mathcal D_{-k}
    \right\}
    =
    \o_p\left(n^{-1/2}\right),
    \E\left\{
        d^2\left(\widehat g_{-k}(\bm X),g(\bm X)\right)
        \mid \mathcal D_{-k}
    \right\}
    =
    \o_p\left(n^{-1/2}\right),
\]
where \((\bm X,Y)\) is an independent copy of the observation.
\end{itemize}

\begin{theorem}\label{THM:crossfit_H0}
{Under $H_0$, suppose that Assumptions (A1), (B0), and (B1) hold and $\E d^4(Y,m(\bm X))>0$. Then, as $n\to\infty$}, \begin{equation*}
\sqrt{n }\widetilde T_{n,K}/\widehat\sigma_{n,K}\xrightarrow{d}\mathcal N(0,1).
\end{equation*}
\end{theorem}

%For fixed alternatives, we replace (B2) by the following weaker consistency condition.
{For fixed alternatives, we impose the following weaker consistency condition.}
\begin{itemize}
\item[(B2')]  For \(k=1,\ldots,K\), the cross-fitted  estimators  {satisfy}
\[
    \E\left\{
        d^2\left(\widehat m_{-k}(\bm X),m(\bm X)\right)
        \mid \mathcal D_{-k}
    \right\}
    =
    \o_p\left(1\right),
    \E\left\{
        d^2\left(\widehat g_{-k}(\bm X),g(\bm X)\right)
        \mid \mathcal D_{-k}
    \right\}
    =
    \o_p\left(1\right),
\]
where \((\bm X,Y)\) is an independent copy of the observation.
\end{itemize}

\begin{theorem}\label{THM:crossfit_H1}
Let $\Delta
=
\E\left\{
d^2(Y,m(\bm X))-d^2(Y,g(\bm X))
\right\}.$
Suppose that Assumptions (B0) and  (B2') hold, $ \E d^4(Y,m(\bm X))>0$,  and   {$ \Delta<0$},   as $n\to\infty$,
\[
\sqrt{n }\widetilde T_{n,K}/\widehat\sigma_{n,K}\xrightarrow{\P}-\infty.
\]
\end{theorem}

% \begin{itemize}
%     \item Cross-fitting
%     \item Introduce $\bm X$-dependent weights (optional)
%     \begin{equation*}%\label{eq:Tnk}
% \breve T_{n}=\frac{1} {n_2}\sum_{i\in \mathcal D_2}w(\bm X_i)\left\{\varphi_{i }d^2(Y_i,\widehat m_{\mathcal D_1}(\bm X_i)  )-\psi_{i }d^2(Y_i,\widehat g_{\mathcal D_1}(\bm X_i)  )\right\}.
% \end{equation*}
% \end{itemize}
Although both the single-split and cross-fitted tests have asymptotic power one
under fixed alternatives, the following theorem compares their power under the local
alternatives  $H_{1n}:\sqrt n \Delta_n\to -\kappa$ for some \(\kappa>0\), where
\begin{align}
    \label{DEF:Delta_n}
    \Delta_n
    =
    \E \left\{d^2(Y,m (\bm X))-d^2(Y,g_n(\bm X))\right\}.
\end{align}
  Here, \(g_n(\cdot)\)
denotes the restricted  population target associated with $n$.
\begin{theorem}\label{THM:local_power_gain}
 Suppose that Assumptions (A0)-(A2), (B0)-(B1) and $ \E d^4(Y,m(\bm X))>0$ hold, with the analogues of (A1), (A2) and (B1) holding uniformly along the local alternative sequence.
 As $n\to\infty$, \begin{align*}
     \frac{\sqrt{n_2}\widetilde T_n}{\widehat\sigma_n}
    \xrightarrow{d}
    \mathcal N\left(
        -\frac{\sqrt{1-\rho}\,\kappa}{\sigma},
        1
    \right),\quad \frac{\sqrt n \widetilde T_{n,K}}{\widehat\sigma_{n,K}}
    \xrightarrow{d}
    \mathcal N\left(
        -\frac{\kappa}{\sigma},
        1
    \right).
 \end{align*}
 Therefore, 
for any significance level $\alpha\in(0,1)$,
 \begin{align*}
     \lim_{n\to\infty}\P\left( \frac{\sqrt{n }\widetilde T_{n,K}}{\widehat\sigma_{n,K}}<z_{\alpha}  \right)>\lim_{n\to\infty}\P\left( \frac{\sqrt{n_2}\widetilde T_n}{\widehat\sigma_n}<z_{\alpha}  \right)>\alpha .
 \end{align*}
\end{theorem}
Theorem~\ref{THM:local_power_gain} rigorously quantifies the power gain from
cross-fitting. Specifically, under local alternatives, the cross-fitted MATCH procedure attains a strictly larger limiting power than the
single-split version. 
Intuitively, cross-fitting increases the effective sample size used for constructing test statistics in the MATCH procedure from $n_2$ to $n$, since every observation serves as a held-out evaluation point in exactly one fold.  Consequently, the local signal is amplified.  
The finite-sample evidence in Section~\ref{sec:simulation} further supports
this theoretical comparison.

\subsection{Repeated cross-fitted MATCH procedures}
To improve the stability of the cross-fitted MATCH procedure, we consider a repeated
cross-fitting implementation. Specifically, for some fixed $J>0$, we randomly shuffle the sample and
perform the (K)-fold cross-fitted MATCH procedure \(J\) times. Each repetition gives an asymptotically valid
left-sided \(p\)-value $p_j=\Phi(\sqrt{n}\widetilde T_{n,K}^{(j)}/\widehat\sigma_{n,K}^{(j)})$, $j=1,\ldots,J$, for the null hypothesis, where $\widetilde T_{n,K}^{(j)}= T_{n,K}^{(j)}-\max\left\{b_{n,K}^{(j)},0\right\}$, \begin{align*}
    &  T_{n,K}^{(j)}
    =
    \frac1n
    \sum_{k=1}^K
    \sum_{i\in\mathcal I_k^{(j)}}
    \left\{
        \varphi_i^{(j)} d^2(Y_i,\widehat m_{-k}^{(j)}(\bm X_i))
        -
        \psi_i^{(j)} d^2(Y_i,\widehat g_{-k}^{(j)}(\bm X_i))
    \right\},\\
    &  b_{n,K}^{(j)}
    =
    \frac1n
    \sum_{k=1}^K
    \sum_{i\in\mathcal I_k^{(j)}}
    \left\{
       d^2(Y_i,\widehat m_{-k}^{(j)}(\bm X_i))
        -
       d^2(Y_i,\widehat g_{-k}^{(j)}(\bm X_i))
    \right\},\\
    &\widehat\sigma_{n,K}^{{(j)}}= \left\{\frac1n
    \sum_{k=1}^K
    \sum_{i\in\mathcal I_k^{(j)}}
    d^4(Y_i,\widehat m_{-k}^{(j)}(\bm X_i))\right\}^{1/2}.
\end{align*} %Since these  \(p\)-values are computed from the same data and are therefore dependent, we combine them using several \(p\)-value merging rules that are designed for, or robust to, dependent \(p\)-values.
Here, the  multipliers $ \left(\varphi_i^{(j)},\psi_i^{(j)}\right)_{i=1}^{n}$ are
generated independently across repetitions $j=1,\ldots,J$   and independently of the data.
%The random fold partitions are also generated independently across repetitions.
Then, we  {aggregate the resulting \(p\)-values using some \(p\)-value merging rule}; see for instance  \cite{liu2020cauchy,vovk2020combining,vovk2021evalues}. Further details are provided in Section~\ref{SEC:p_merging}.

Among these methods, we recommend the Cauchy combination approach as the default choice, and summarize   the resulting procedure in Algorithm~\ref{alg:repeated_cross_fitting}.

\begin{algorithm}[h!]
\caption{Repeated cross-fitted MATCH procedure}
\label{alg:repeated_cross_fitting}
\begin{algorithmic}[1]
\Require
Data $\mathcal D_n=\{(Y_i,\bm X_i)\}_{i=1}^n$;
restricted objective \(G\) defining the target \(g\);
number of folds $K$;
number of repetitions $J$.

\Ensure

The Cauchy-aggregated MATCH $p$-value {$p_{\mathrm{cauchy}}$}.

\For{$j=1,\ldots,J$}
    \State Randomly shuffle the observations and partition
    $\{1,\ldots,n\}$ into $K$ disjoint folds
    $\mathcal I_1^{(j)},\ldots,\mathcal I_K^{(j)}$. Let
        $\mathcal D_{-k}^{(j)}
        =
        \mathcal D_n\setminus
        \{(Y_i,\bm X_i):i\in\mathcal I_k^{(j)}\}$.

    \For{$k=1,\ldots,K$}

        \State
        Using $\mathcal D_{-k}^{(j)}$, estimate the unrestricted conditional \Frechet{} mean \(m\) and the restricted target \(g\), yielding
        $\widehat m_{-k}^{(j)}$ and
        $\widehat g_{-k}^{(j)}$.

        \State Evaluate the fold-specific contributions on
        $\mathcal I_k^{(j)}$ using
        $\widehat m_{-k}^{(j)}$ and
        $\widehat g_{-k}^{(j)}$.
    \EndFor

    \State Aggregate the $K$ fold-specific contributions to obtain
    the cross-fitted statistic
    $\sqrt{n}\widetilde T_{n,K}^{(j)}/\widehat\sigma_{n,K}^{(j)}$ and compute the left-sided $p$-value $
        p_j
        =
        \Phi\left(
             {\sqrt n\,\widetilde T_{n,K}^{(j)}}/
                 {\widehat\sigma_{n,K}^{(j)}}
        \right). $
\EndFor

\State Compute the Cauchy-aggregated $p$-value
\[
    p_{\mathrm{cauchy}}
    =
    \frac12-\frac{1}{\pi}\arctan\left( \frac{1}{J}
    \sum_{j=1}^J
    \tan\left\{
        \pi\left(\frac12-p_j\right)
    \right\}\right).
\]

\State \Return $p_{\mathrm{cauchy}}$.

\end{algorithmic}
\end{algorithm}

The following theorem establishes the asymptotic validity of the proposed Cauchy-aggregated testing procedure under the null hypothesis and its consistency against alternatives whose signal dominates the $n^{-1/2}$ scale.

\begin{theorem}\label{THM:repeated}
Consider Algorithm \ref{alg:repeated_cross_fitting} and suppose that Assumptions (A1), (B0), (B1) and $ \E d^4(Y,m(\bm X))>0$ hold. For any $\alpha\in(0,1)$,
under the null,
\[
    \lim_{n\to\infty}
    \mathbb P\left(p_{\mathrm{cauchy}}\leq\alpha\right)
    =
    \alpha.
\]
Under   a sequence of alternatives such that $\sqrt n\Delta_n\to -\infty$,  with $\Delta_n$   defined in \eqref{DEF:Delta_n},
\[
    \lim_{n\to\infty}
    \mathbb P\left(p_{\mathrm{cauchy}}\leq\alpha\right)
    =
    1.
\]
\end{theorem}

\section{Simulation Studies}\label{sec:simulation}

We evaluate the finite-sample performance of the proposed significance tests for
Fréchet regression on three non-Euclidean response spaces:  the
one-dimensional Wasserstein distribution space,  the space of symmetric
positive-definite matrices, and the unit sphere.
In all simulations, the covariates $\bm X=(X_1,X_2)^\top$ are generated as
$X_1,X_2\overset{\mathrm{i.i.d.}}{\sim} \operatorname{Unif}[-1,1]$.
We consider three testing problems: global significance, partial significance,
and global \Frechet{} specification testing.  For the partial significance test, we write
\(\bm X=(Z,W)^\top\), with \(Z=X_1\) and \(W=X_2\).

For each setting, we compare the original sample-splitting statistic and its
\(K\)-fold cross-fitting version.  The original statistic uses one random split
\(\mathcal D=\mathcal D_1\cup \mathcal D_2\), with
\(|\mathcal D_1|=\lfloor n/2\rfloor\), where the nuisance functions are
estimated on \(\mathcal D_1\) and evaluated on \(\mathcal D_2\).  The
cross-fitting version uses \(K=5\) folds.  For each fold \(\mathcal I_k\), the
estimators \(\widehat m_{-k}\) and \(\widehat g_{-k}\) are trained on
\(\mathcal D_{-k}=\{(\bm X_i,Y_i):i\notin \mathcal I_k\}\) and evaluated
on observations in \(\mathcal I_k\). The multipliers $\varphi_i,\psi_i\overset{\mathrm{i.i.d.}}{\sim}N(1,1/2)$ are independent of the data.  We set the significance level \(\alpha=0.05\).

% Throughout the simulations, we use the two-multiplier statistic
% \[
%     T_n
%     =
%     \frac{1}{n_2}
%     \sum_{i\in\mathcal D_2}
%     \left[
%         \varphi_i d^2\{Y_i,\widehat m(\bm X_i)\}
%         -
%         \psi_i d^2\{Y_i,\widehat g(\bm X_i)\}
%     \right],
% \]
% where
% \[
%     \varphi_i,\psi_i\overset{\mathrm{i.i.d.}}{\sim}N(1,1/2).
% \]
% The corresponding \(K\)-fold statistic is
% \[
%     T_{n,K}
%     =
%     \frac1n
%     \sum_{k=1}^K
%     \sum_{i\in\mathcal I_k}
%     \left[
%         \varphi_i d^2\{Y_i,\widehat m_{-k}(\bm X_i)\}
%         -
%         \psi_i d^2\{Y_i,\widehat g_{-k}(\bm X_i)\}
%     \right],
% \]
% where the multipliers are independent of the data and independent across
% observations.  In both cases, the variance estimator is
% \[
%     \widehat\sigma^2
%     =
%     \frac1{n_\star}
%     \sum
%     d^4\{Y_i,\widehat m(\bm X_i)\},
% \]
% where \(n_\star=n_2\) for the original sample split and \(n_\star=n\) for the
% cross-fitting version.  The null hypothesis is rejected when the corresponding studentized statistic is smaller than \(z_\alpha\), the lower \(\alpha\)-quantile of the standard Gaussian distribution.  We set \(\alpha=0.05\).

The unrestricted conditional \Frechet{} mean is estimated by local   Fréchet
regression with the Epanechnikov product kernel
\[
    K(\bm u)
    =
    \prod_{j=1}^q
    \frac34(1-u_j^2)\mathbf 1(|u_j|\le 1),
\]
where \(q\) is the dimension of the covariate used in the
estimator.
The bandwidth is set to
$h=n_{\mathrm{train}}^{-1/(4+q)}$,
where \(n_{\mathrm{train}}\) is the training sample size.
% for the corresponding
% split or fold, and   Thus \(q=2\) for estimating the full conditional target
% \(m(\bm x)\), whereas \(q=1\) for estimating the partial target \(f(z)\).
For the global \Frechet{} specification test, the restricted estimator is computed using the empirical
linear weight
\(
    \widehat s(\bm X_i,\bm x)
    =
    1+
    (\bm X_i-\widehat{\bm \mu} )^\top
    \widehat{\mathbf \Sigma}^{-1}
    (\bm x-\widehat{\bm \mu} ).
\)
The empirical rejection probabilities are computed over \(500\) Monte Carlo
replications for each sample size \(n\in\{200,400\}\).

Additional simulations are provided in Section \ref{sec:additional-sim} of the Supplementary Material, covering alternative $p$-value merging rules, sensitivity to the bandwidth, the number of cross-fitting folds $K$, and the choice of multipliers, as well as the asymmetric randomization approach in Remark \ref{remark2}.

\subsection{Distributional responses under the Wasserstein metric}\label{sec:simulation-dist}

We first consider one-dimensional distributional responses, where \(\Omega\) is
the set of distributions on \([0,1]\) with finite second moment, equipped with
the  Wasserstein distance.
% In one dimension,
% \[
%     W_2^2(\nu_1,\nu_2)
%     =
%     \int_0^1
%     \{Q_{\nu_1}(t)-Q_{\nu_2}(t)\}^2\,dt,
% \]
% where \(Q_\nu\) denotes the quantile function of \(\nu\).
  Let
\(
    b(t)=t(1-t).
\)
The response distribution is generated through its quantile function:
\[
    Q_Y(t)=t+\xi(\bm X,\varepsilon)b(t),
    \qquad 0\le t\le 1,
\]
where
$\varepsilon\sim\operatorname{Unif}[-0.1,0.1]$
is independent of \(\bm X\), and
\[
    \xi(\bm X,\varepsilon)
    =
    \theta_1X_1+\theta_2X_2
    +\beta_1(X_1^2-1/3)
    +\beta_2\{\cos(X_2)-\sin(1)\}
    +\varepsilon.
\]
For the parameter grids below, \(|\xi(\bm X,\varepsilon)|<1\), and hence
\(
    Q_Y'(t)
    =
    1+\xi(\bm X,\varepsilon)(1-2t)>0.
\)
Therefore \(Q_Y(\cdot)\) is a valid quantile function.
\begin{comment}
Since all responses share the
same basis \(b(t)\),
\[
    W_2^2(Y_1,Y_2)
    =
    \{A_1-A_2\}^2\int_0^1 b^2(t)\,dt
    =
    \frac{(A_1-A_2)^2}{30}.
\]
Consequently, the local Fréchet estimator can be computed by applying local
weighted averaging to the scalar amplitude \(A\).
\end{comment}

%The coefficient settings are parallel to those in the spherical simulation.
\begin{itemize}
    \item[(a)] For the global significance test,
 $ (\theta_1,\beta_1,\theta_2,\beta_2)
    =
    a(1,1,1,1)$, with $
    a\in\\\{0,0.01,0.02,0.03,0.04,0.05\}$.
    \item[(b)] For the partial significance test,
 $\theta_1=0.45$,
    $\beta_1=0$,
    $\theta_2=0$,
    $\beta_2=a$, with $
    a\in\{0,0.1,0.2,0.3,0.4,0.5\}$.
    \item[(c)] For the global \Frechet{} specification test,
 $\theta_1=0.35$,
    $\theta_2=0.35$,
    $\beta_1=a$,
   $\beta_2=0$, with $
    a\in\{0,0.05,0.1,0.15,0.2,0.25\}$.
\end{itemize}
The parameter \(a\) controls the departure from the corresponding null
hypothesis.

\subsection{SPD responses under the log-Euclidean metric}

We next consider responses in the space \(\mathcal S_{++}^3\) of
\(3\times 3\) symmetric positive-definite matrices under the
log-Euclidean metric.
% \[
%     d_{\mathrm{LE}}(A,B)
%     =
%     \|\log A-\log B\|_F.
% \]
Let \(\mathbf B_0=0.2\mathbf I_3\), and define the following Frobenius-orthonormal symmetric
directions:
\[
    \mathbf H_1=\frac1{\sqrt2}\operatorname{diag}(1,-1,0),
    \quad
    \mathbf H_2=\frac1{\sqrt2}(\mathbf E_{12}+\mathbf E_{21}),
\]
\[
    \mathbf H_3=\frac1{\sqrt6}\operatorname{diag}(1,1,-2),
    \quad
    \mathbf H_4=\frac1{\sqrt2}(\mathbf E_{13}+\mathbf E_{31}),
\]
where \(\mathbf E_{ij}\) has a one in position \((i,j)\) and zeros elsewhere.  We
generate $Y=\exp(\mathbf S_Y)$, where
\[
    \mathbf S_Y
    =
    \mathbf B_0
    +
    \theta_1X_1\mathbf H_1
    +
    \theta_2X_2\mathbf H_2
    +
    \beta_1(X_1^2-1/3)\mathbf H_3
    +
    \beta_2\{\cos(X_2)-\sin(1)\}\mathbf H_4
    +
    \mathbf E.
\]
Here, \(\mathbf E\) is a mean-zero symmetric noise matrix whose independent
upper-triangular entries are drawn from \(N(0,\sigma_E^2)\), with
\(\sigma_E=0.05\).
% Since \(\mathbf S_Y\) is symmetric, \(Y=\exp(S_Y)\) is symmetric
% positive definite.
% Moreover, under the log-Euclidean metric,
% \[
%     d_{\mathrm{LE}}^2(Y_1,Y_2)
%     =
%     \|S_{Y_1}-S_{Y_2}\|_F^2,
% \]
% so Fréchet regression reduces to ordinary weighted averaging in the log-domain.
Although normal log-domain noise yields unbounded log-Euclidean SPD responses,   the finite-diameter condition is primarily a technical assumption for theoretical analysis rather than a strict requirement for practical effectiveness. Similar normal noise settings can be found in \cite{IaoZhouMuller2025}.
\begin{itemize}
    \item[(a)] For the global significance test,
 $ (\theta_1,\beta_1,\theta_2,\beta_2)
    =
    a(1,1,1,1)$, with $
    a\in\\\{0,0.03,0.06,0.09,0.12,0.15\}$.
    \item[(b)] For the partial significance test,
 $\theta_1=0.45$,
    $\beta_1=0$,
    $\theta_2=0$,
    $\beta_2=a$, with $
    a\in\{0,0.2,0.4,0.6,0.8,1\}$.
    \item[(c)] For the global \Frechet{} specification test,
 $\theta_1=0.35$,
    $\theta_2=0.35$,
    $\beta_1=a$,
    $\beta_2=0$, with $
    a\in\{0,0.1,0.2,0.3,0.4,0.5\}$.
\end{itemize}

\subsection{Spherical responses}

We finally consider responses on the unit sphere \(S^2\) equipped with the
geodesic distance \(d_g\).  Let $ \bm\mu=(1,0,0)^\top$, $\bm e_1=(0,1,0)^\top$, and $\bm e_2=(0,0,1)^\top$,
where \(\bm e_1,\bm e_2\) form an orthonormal basis of \(T_\mu S^2\).  For a given
\(\bm X\), define
\[
    \bm \xi(\bm X)
    =
    \{\theta_1X_1+\beta_1(X_1^2-1/3)\}\bm e_1
    +
    \left[ \theta_2X_2+\beta_2\{\cos(X_2)-\sin(1) \} \right] \bm  e_2,
\]
and set
$\bm \eta(\bm X)=\operatorname{Exp}_{\bm \mu}\{\bm \xi(\bm X)\}$.
The response is generated from a small isotropic tangent perturbation around
\(\bm \eta(\bm X)\): $Y=\operatorname{Exp}_{\bm \eta(\bm X)}(\bm \varepsilon)$, where  $\bm \varepsilon=\sigma_Y(N_1 \bm b_1+N_2 \bm b_2)$,
  \(N_1,N_2\overset{\mathrm{i.i.d.}}{\sim}N(0,1)\),
\(\sigma_Y=0.25\), and \(\bm b_1,\bm b_2\) are an orthonormal basis of
\(T_{\bm \eta(\bm X)}S^2\). The distribution of \(\bm\varepsilon\) does not depend on the
particular choice of this basis, since the perturbation is isotropic in the
tangent space.

\begin{itemize}
    \item[(a)] For the global significance test,
 $ (\theta_1,\beta_1,\theta_2,\beta_2)
    =
    a(1,1,1,1)$, with $
    a\in\\\{0,0.08,0.16,0.24,0.32,0.4\}$.
    \item[(b)] For the partial significance test,
 $\theta_1=0.45$,
    $\beta_1=0$,
    $\theta_2=0$,
    $\beta_2=a$, with $
    a\in\{0,0.4,0.8,1.2,1.6,2\}$.
    \item[(c)] For the global \Frechet{} specification test,
 $\theta_1=0.35$,
    $\theta_2=0.35$,
    $\beta_1=a$,
    $\beta_2=0$, with $
    a\in\{0,0.3,0.6,0.9,1.2,1.5\}$.
\end{itemize}

% The three alternatives are generated as follows.  For the global significance
% test,
% \[
%     (\theta_1,\beta_1,\theta_2,\beta_2)
%     =
%     a(1,1,1,1),
%     \qquad
%     a\in\{0,0.40/6,2(0.40/6),\ldots,0.40\}.
% \]
% For the partial significance test,
% \[
%     \theta_1=0.45,\qquad
%     \beta_1=0,\qquad
%     \theta_2=0,\qquad
%     \beta_2=a,
%     \qquad
%     a\in\{0,2/6,2(2/6),\ldots,2\}.
% \]
% For the global linearity test,
% \[
%     \theta_1=0.35,\qquad
%     \theta_2=0.35,\qquad
%     \beta_1=a,\qquad
%     \beta_2=0,
%     \qquad
%     a\in\{0,1.5/6,2(1.5/6),\ldots,1.5\}.
% \]

\subsection{Results}
\begin{figure}[h!]
    \centering
    \includegraphics[width=0.32\linewidth]{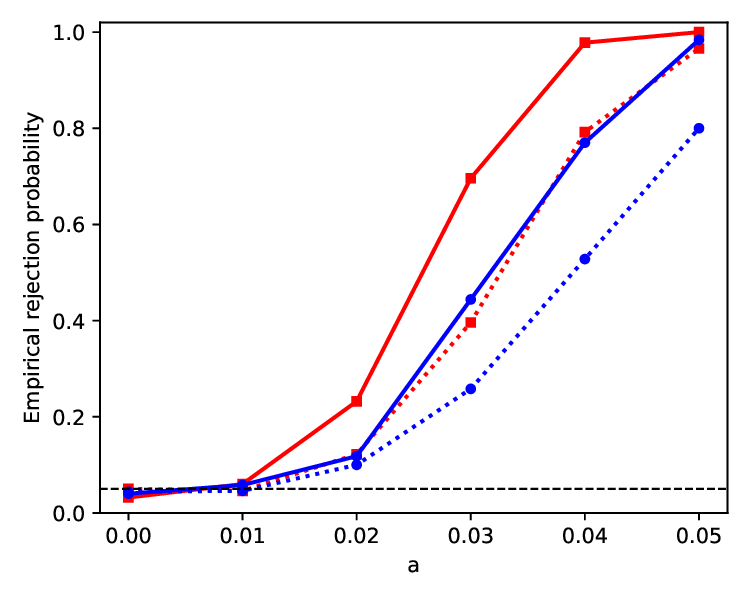}
    \includegraphics[width=0.32\linewidth]{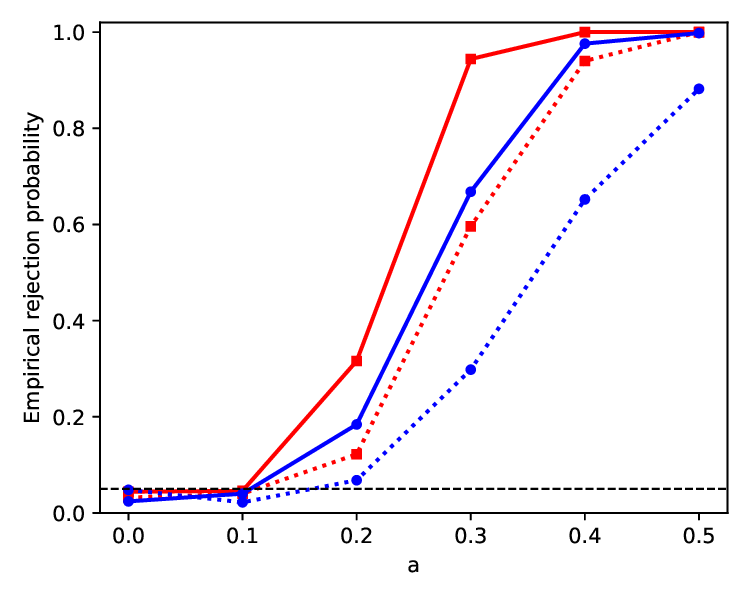}
    \includegraphics[width=0.32\linewidth]{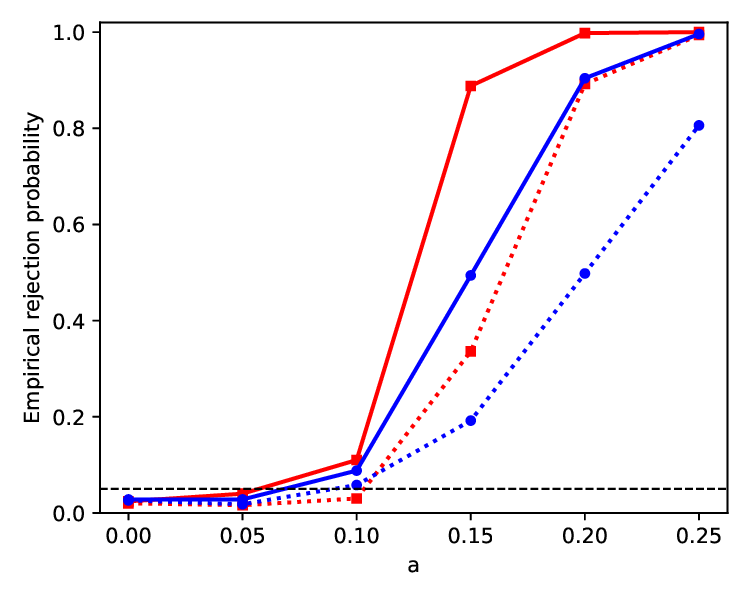}

    \caption{Empirical power curves for distributional responses. Panels from left to right show
the global, partial, and global \Frechet{} specification tests. Dotted and solid lines correspond to
\(n=200\) and \(n=400\), respectively. Blue circles denote \(\widetilde T_n\), and red
squares denote \(\widetilde T_{n,K}\). The dashed horizontal line marks \(\alpha=0.05\).}
    \label{fig:distribution}
\end{figure}

\begin{figure}[h!]
    \centering
\includegraphics[width=0.32\linewidth]{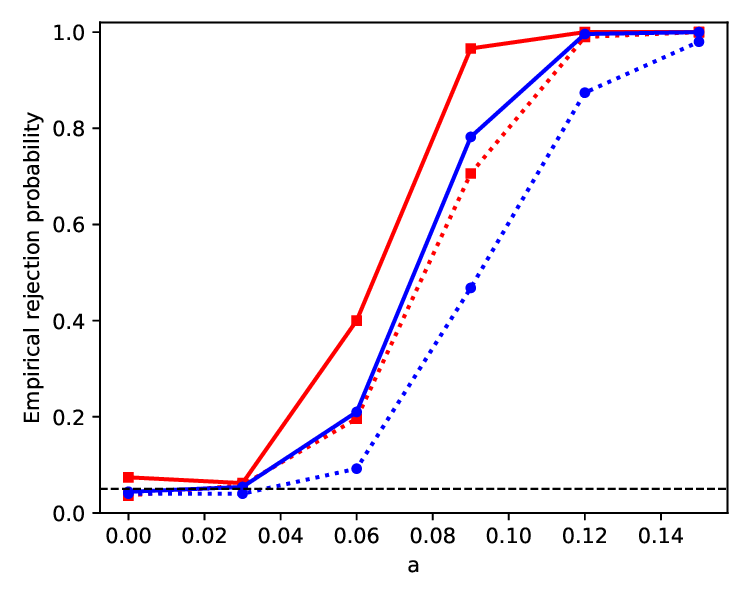}
    \includegraphics[width=0.32\linewidth]{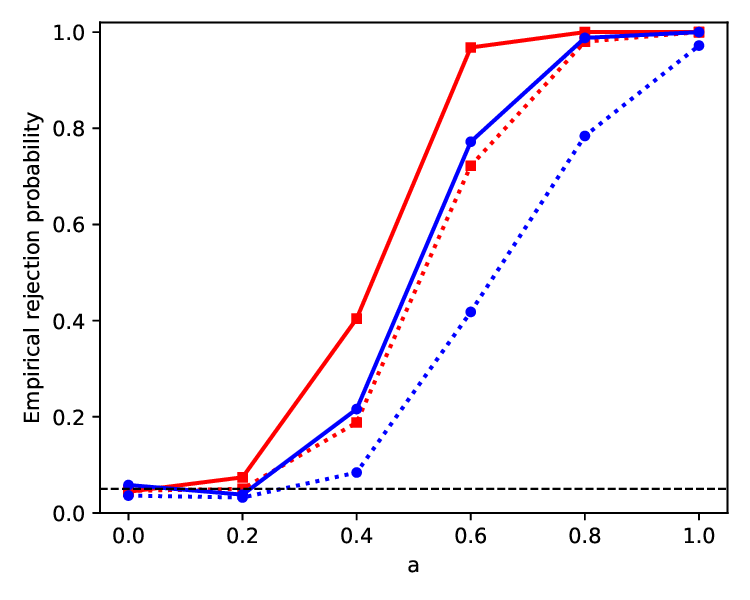}
    \includegraphics[width=0.32\linewidth]{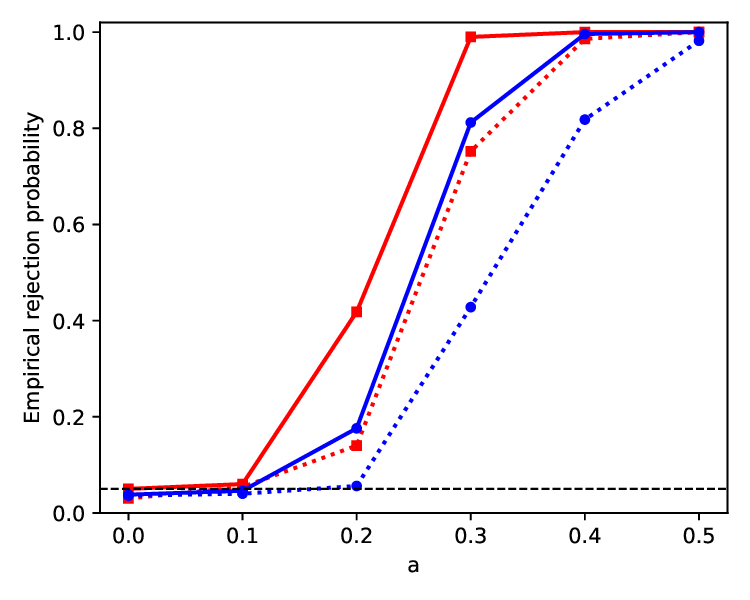}

    \caption{Empirical power curves for SPD responses. Panels from left to right show
the global, partial, and global \Frechet{} specification tests. Dotted and solid lines correspond to
\(n=200\) and \(n=400\), respectively. Blue circles denote \(\widetilde T_n\), and red
squares denote \(\widetilde T_{n,K}\). The dashed horizontal line marks \(\alpha=0.05\).}
    \label{fig:spd}
\end{figure}

\begin{figure}[h!]
    \centering
    \includegraphics[width=0.32\linewidth]{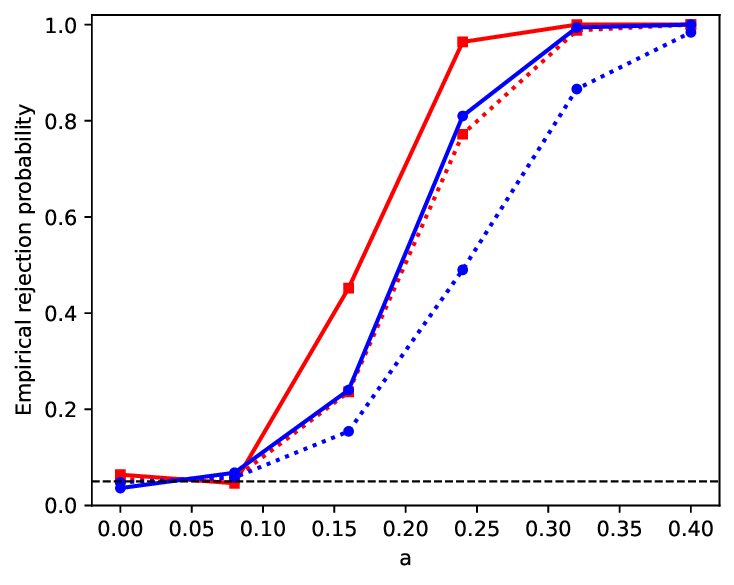}
    \includegraphics[width=0.32\linewidth]{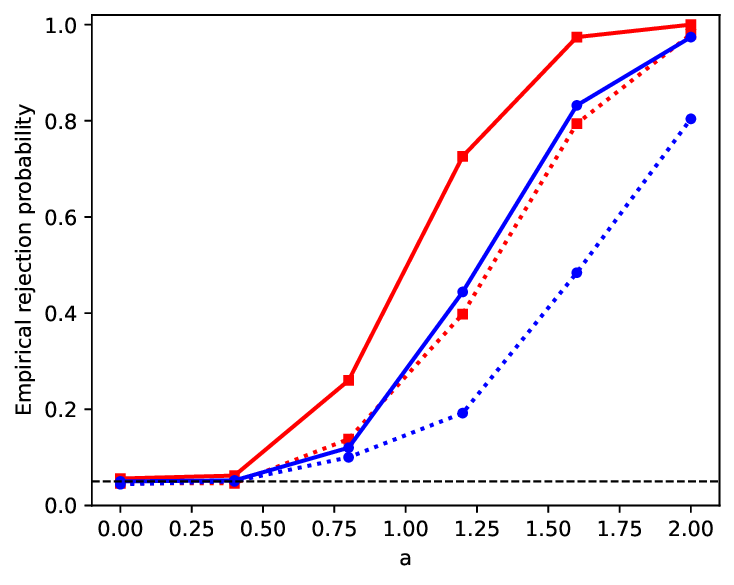}
    \includegraphics[width=0.32\linewidth]{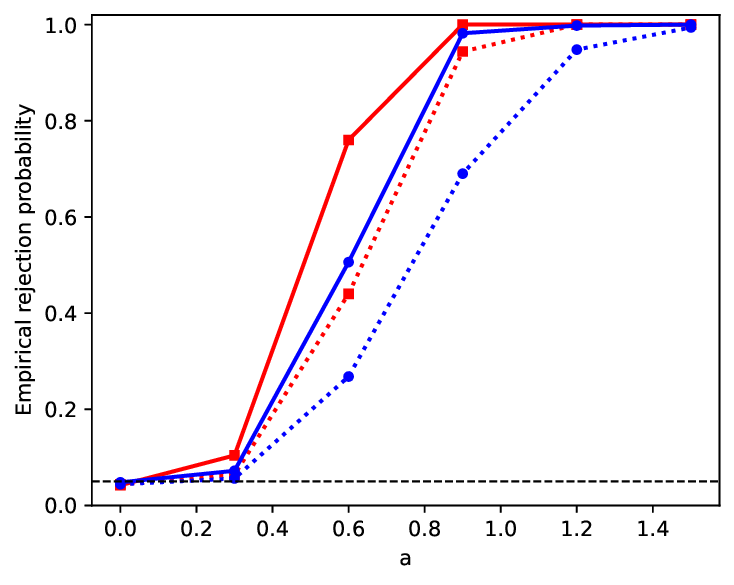}

    \caption{Empirical power curves for spherical responses. Panels from left to right show
the global, partial, and global \Frechet{} specification tests. Dotted and solid lines correspond to
\(n=200\) and \(n=400\), respectively. Blue circles denote \(\widetilde T_n\), and red
squares denote \(\widetilde T_{n,K}\). The dashed horizontal line marks \(\alpha=0.05\).}
    \label{fig:sphere}
\end{figure}

For each response space and each testing problem, empirical power curves in Figures \ref{fig:distribution}-\ref{fig:sphere} are
obtained by plotting the rejection probability against \(a\).  Each figure
contains four curves:    the single-split   statistic $\widetilde T_n$  with
\(n=200,400\), the cross-fitted  statistic $\widetilde T_{n,K}$  with
\(n=200,400\),   together with a
horizontal reference line at \(\alpha=0.05\).
At a fixed sample size, the empirical power increases   with $a$. The cross-fitted test generally has higher empirical power than the single-split test. Table~\ref{tab:empirical_sizes} reports the empirical sizes for all response spaces and testing problems, which  are generally close to the nominal level $0.05$, with
some minor finite-sample deviations.

% \begin{table}[h!]
% \centering
% \caption{Empirical sizes  for distributional responses under the   null hypotheses. }
% \label{tab:wasserstein_size}
% \begin{tabular}{lcccc}
% \toprule
% & \multicolumn{2}{c}{$n=200$}
% & \multicolumn{2}{c}{$n=400$} \\
% \cmidrule(lr){2-3}\cmidrule(lr){4-5}
% Testing problem
% & $\widetilde T_n$
% & $\widetilde T_{n,K}$
% & $\widetilde T_n$
% & $\widetilde T_{n,K}$ \\
% \midrule
% Global significance  & 0.044 & 0.050 & 0.040 & 0.032 \\
% Partial significance & 0.048 & 0.032 & 0.024 & 0.044 \\
% Linearity            & 0.026 & 0.020 & 0.028 & 0.024 \\
% \bottomrule
% \end{tabular}
% \end{table}

% \begin{table}[h!]
% \centering
% \caption{Empirical sizes  for SPD responses under the   null hypotheses.  }
% \label{tab:spd_size}
% \begin{tabular}{lcccc}
% \toprule
% & \multicolumn{2}{c}{$n=200$}
% & \multicolumn{2}{c}{$n=400$} \\
% \cmidrule(lr){2-3}\cmidrule(lr){4-5}
% Testing problem
% & $\widetilde T_n$
% & $\widetilde T_{n,K}$
% & $\widetilde T_n$
% & $\widetilde T_{n,K}$ \\
% \midrule
% Global significance  & 0.040 & 0.036 & 0.044 & 0.074 \\
% Partial significance & 0.036 & 0.046 & 0.058 & 0.044 \\
% Linearity            & 0.036 & 0.030 & 0.038 & 0.050 \\
% \bottomrule
% \end{tabular}
% \end{table}

% \begin{table}[h!]
% \centering
% \caption{Empirical sizes for spherical responses under the  null hypotheses. }
% \label{tab:sphere_size}
% \begin{tabular}{lcccc}
% \toprule
% & \multicolumn{2}{c}{$n=200$}
% & \multicolumn{2}{c}{$n=400$} \\
% \cmidrule(lr){2-3}\cmidrule(lr){4-5}
% Testing problem
% & $\widetilde T_n$
% & $\widetilde T_{n,K}$
% & $\widetilde T_n$
% & $\widetilde T_{n,K}$ \\
% \midrule
% Global significance  & 0.048 & 0.054 & 0.036 & 0.064 \\
% Partial significance & 0.044 & 0.046 & 0.050 & 0.056 \\
% Linearity            & 0.044 & 0.044 & 0.048 & 0.042 \\
% \bottomrule
% \end{tabular}
% \end{table}

\begin{table}[h!]
\centering
\caption{Empirical sizes of the proposed tests for distributional, SPD, and spherical responses under the null hypotheses.}
\label{tab:empirical_sizes}
\begin{tabular}{llcccc}
\toprule
&& \multicolumn{2}{c}{$n=200$}
& \multicolumn{2}{c}{$n=400$} \\
\cmidrule(lr){3-4}\cmidrule(lr){5-6}
Response space
& Testing problem
& single-split
& cross-fitted
& single-split
& cross-fitted \\
\midrule
Distributional
& Global significance  & 0.044 & 0.050 & 0.040 & 0.032 \\
& Partial significance & 0.048 & 0.032 & 0.024 & 0.044 \\
& Linearity            & 0.026 & 0.020 & 0.028 & 0.024 \\
\addlinespace

SPD
& Global significance  & 0.040 & 0.036 & 0.044 & 0.074 \\
& Partial significance & 0.036 & 0.046 & 0.058 & 0.044 \\
& Linearity            & 0.036 & 0.030 & 0.038 & 0.050 \\
\addlinespace

Spherical
& Global significance  & 0.048 & 0.054 & 0.036 & 0.064 \\
& Partial significance & 0.044 & 0.046 & 0.050 & 0.056 \\
& Linearity            & 0.044 & 0.044 & 0.048 & 0.042 \\
\bottomrule
\end{tabular}
\end{table}

\section{Real data analysis}\label{sec:realdata}

\subsection{Income distribution data}\label{sec:income_data}

We illustrate the proposed testing procedures using county-level household income distributions from the 2020--2024 American Community Survey (ACS) five-year estimates, table B19001, which is available from the U.S. Census Bureau at
\url{https://www.census.gov/data}.

\paragraph{Data processing}
The analysis includes counties in the 50 states and the District of Columbia, while counties with fewer than 500 estimated households are excluded. After preprocessing, the final sample contains $n=3096$ counties.
For each county, the ACS table B19001 reports the estimated numbers of households in 16 ordered income categories, ranging from less than \$10,000 to \$200,000 or more.   Let
\(
    p_{i\ell}
    =
     {N_{i\ell}}
    /{\sum_{r=1}^{16}N_{ir}}\)
    for
   \( \ell=1,\ldots,16\), where $N_{i\ell}$ denotes the number of households in the $\ell$th income category for county $i$.
We construct the county-level household income distribution by
\(
    Y_i
    =
    \sum_{\ell=1}^{16}p_{i\ell}\delta_{\log(1+c_\ell)},
\)
where $c_\ell$ is a representative income value for the $\ell$th category. For the first 15 ordered  categories, we use the corresponding interval midpoints for $c_\ell$, while the last category of ``\$200,000 or more'' is represented by \$250,000. Here,   the $\log(1+\text{income})$ scale is adopted to reduce the strong right skewness of household income.

% For a distribution $Y_i$, let $Q_i(u)$ denote its quantile function on the transformed income scale.
% The squared Wasserstein distance between two county-level distributions is approximated by
% \[
%     W_2^2(Y_i,Y_j)
%     \approx
%     \frac{1}{L}
%     \sum_{r=1}^{L}
%     \left\{
%         Q_i(u_r)-Q_j(u_r)
%     \right\}^2,
% \]
% where $\{u_r\}_{r=1}^{L}$ is an equally spaced grid on $(0,1)$. We use $L=51$ in the analysis.
We consider the following four county-level predictors:
\(
    \bm X
    =
    \left(
        \bm Z^\top, W
    \right)^\top
\), in which
\(
    W=(\text{college share})
\)
and
\(
    \bm Z
    =
    \left(
        \text{unemployment rate},
        \text{median age},
        \log(1+\text{population density})
    \right)^\top.
\)
Here, college share is the proportion of residents aged 25 years or older with a bachelor's degree or higher, unemployment rate is the proportion of unemployed individuals in the civilian labor force, median age measures the age composition of the county, and population density is the total population divided by the county land area in square miles.

%We use $K=5$ folds and repeat the cross-fitting procedure $J=20$ times. Fresh Gaussian multipliers are generated independently across repetitions, and the resulting left-sided $p$-values are aggregated using the Cauchy combination rule. The finite-sample correction $\max\{b_{n,K},0\}$ is applied in every repetition.

\paragraph{Implementation and results}
We consider global significance, partial significance, and global \Frechet{} specification testing problems.  The unrestricted conditional \Frechet{} mean and the reduced conditional \Frechet{} mean are estimated by local \Frechet{} regression using adaptive Epanechnikov kernel weights. For each evaluation point, the bandwidth is determined by the distance to its 100th nearest neighbor in the standardized predictor space. The restricted model in the global \Frechet{} specification test is estimated by global \Frechet{} regression.

As reported in Table~\ref{tab:acs_results}, the global significance test returns a very small \(p\)-value, providing evidence that the conditional \Frechet{} mean income distribution  {is associated with} the four predictors.
The partial significance test also rejects the null, suggesting that college attainment provides additional information about the conditional \Frechet{} mean income distribution after controlling for labor-market conditions, age composition, and population density.

\begin{figure}[h!]
\centering
\includegraphics[width=0.65\textwidth]
{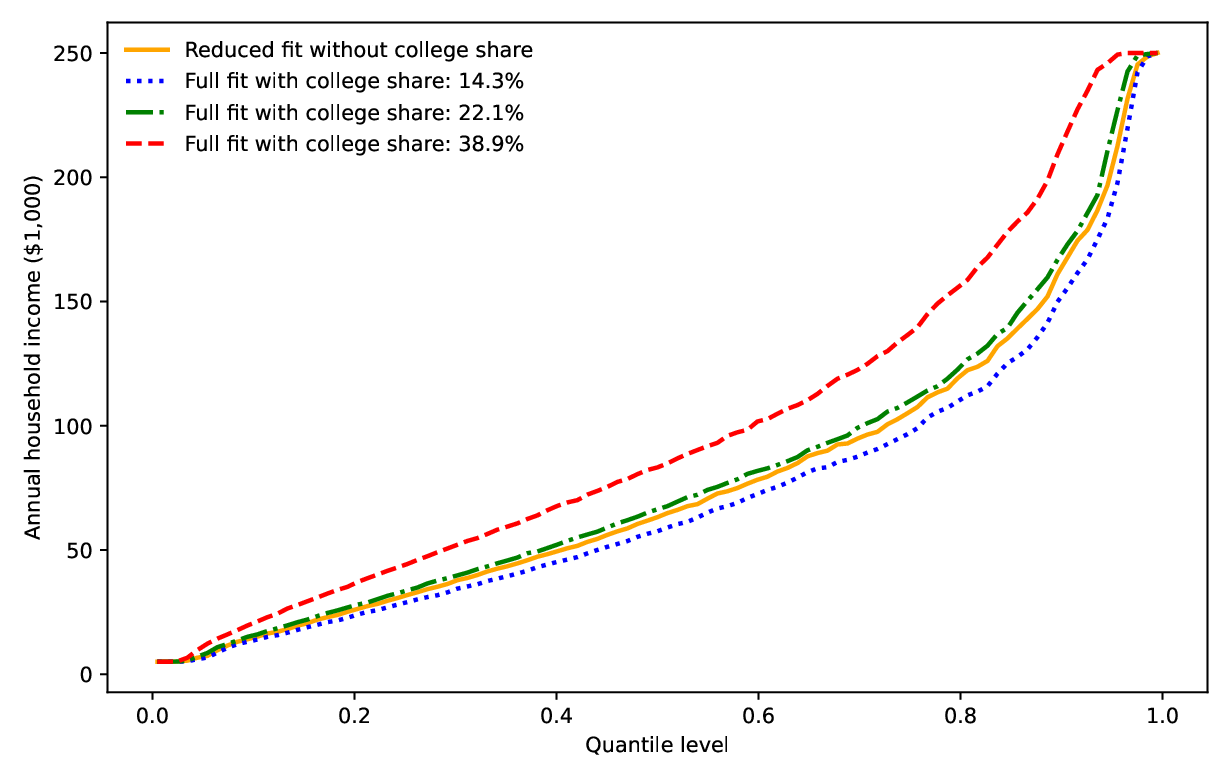}
\caption{Full and reduced fitted conditional \Frechet{} mean income distributions. The reduced fit excludes college share, while the full fits include college share fixed at its 10th, 50th, and 90th sample percentiles. The remaining predictors are fixed at their sample medians. }
\label{fig:acs_partial_fit}
\end{figure}

To visualize the partial effect of college attainment, we fix the components of $\bm Z$ at their sample medians:
\(
    \text{unemployment rate}=0.0443\),
    \(\text{median age}=41.5\),
 and
\(
    \text{population density}=45.8\)
 persons per square mile.
We then compare the reduced fitted distribution based only on $\bm Z$ with the full fitted distributions evaluated at the 10th, 50th, and 90th percentiles of college share, corresponding to approximately $14.3\%$, $22.1\%$, and $38.9\%$, respectively.
Figure~\ref{fig:acs_partial_fit} shows that the fitted income distribution shifts upward as college share increases, which is consistent with the strong rejection obtained from the partial significance test.   %The separation between the full and reduced fitted quantile functions is consistent with the strong rejection obtained from the partial significance test.

\begin{figure}[h!]
\centering
\includegraphics[width=0.49\textwidth]
{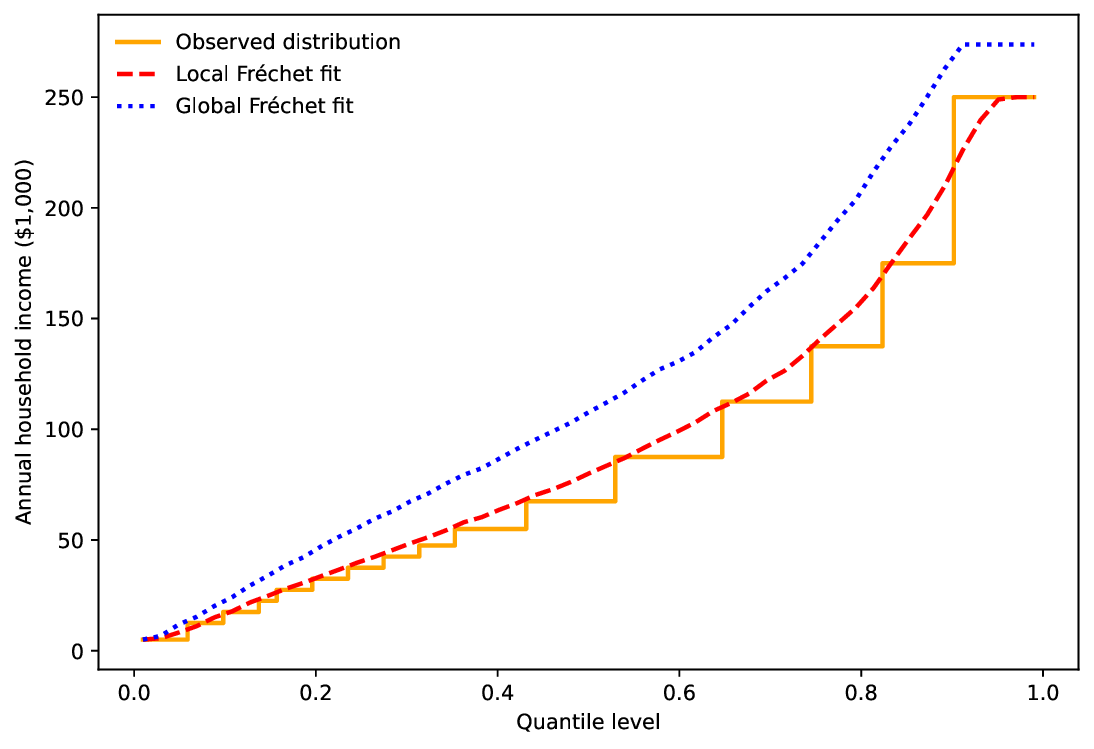}
\includegraphics[width=0.49\textwidth]
{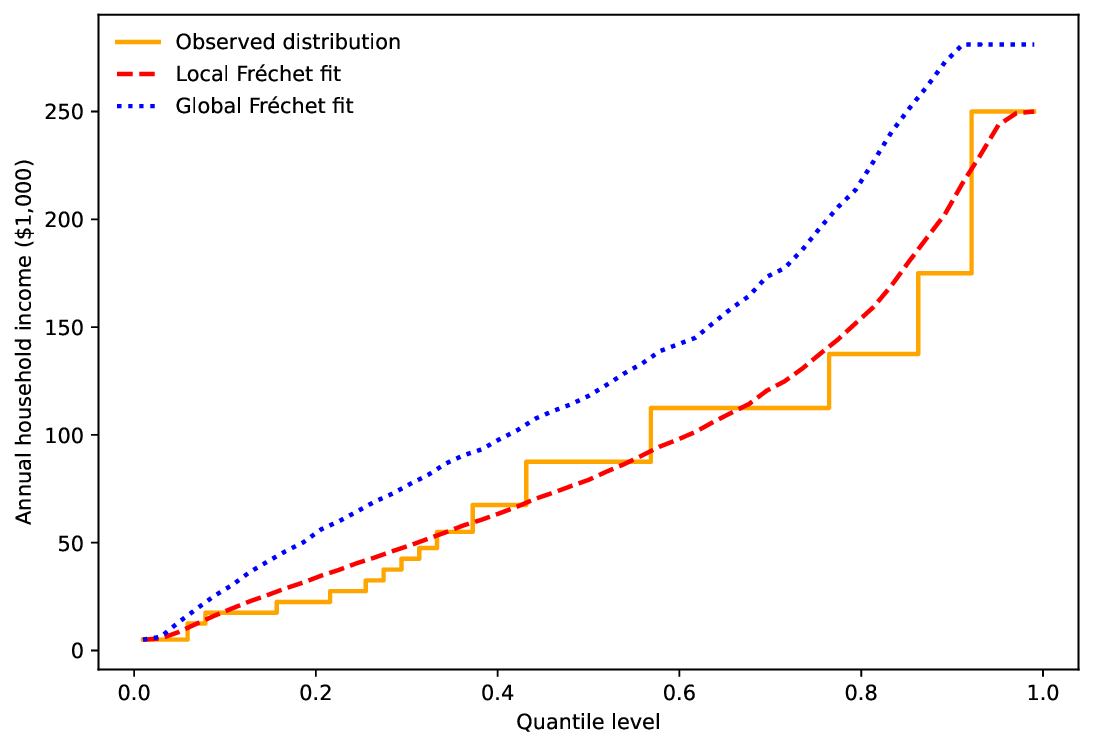}
\caption{Observed and fitted conditional \Frechet{} mean income distributions for two representative counties. (Left panel): Story County, Iowa. (Right panel): Lexington City, Virginia.}
\label{fig:county_fits}
\end{figure}

Finally, the global \Frechet{} specification test rejects the global \Frechet{} regression specification, suggesting that the relationship between the county characteristics and the conditional \Frechet{} mean income distribution is not adequately described by the global linear structure. 
Figure~\ref{fig:county_fits} compares the observed income distributions with the out-of-fold local and global \Frechet{} regression fits for  Story County, Iowa, and Lexington City, Virginia. In both counties, the local estimator follows the observed distribution more closely, particularly in the middle and upper quantiles. For Lexington City, the squared Wasserstein loss decreases from $0.252$ under global \Frechet{} regression to $0.035$ under local \Frechet{} regression, corresponding to an $86.0\%$ reduction. For Story County, the loss decreases from $0.151$ to $0.017$, a reduction of $88.6\%$. These county-level comparisons illustrate the lack of fit underlying the rejection of the global \Frechet{} regression specification.

\begin{table}[h!]
\centering
\caption{The $p$-values of the proposed multiplier-assisted tests for the ACS county-level household income distributions.}
\label{tab:acs_results}
\begin{tabular}{lccc}
\toprule
Testing problem
& Single-split
& Cross-fitted
& Cauchy-aggregated   \\
\midrule
Global significance
& $<1.00 \times10^{-6}$
& $<1.00 \times10^{-6}$
& $<1.00 \times10^{-6}$ \\

Partial significance
& $<1.00 \times10^{-6}$
& $<1.00 \times10^{-6}$
& $<1.00 \times10^{-6}$ \\

 Global specification
& $3.57 \times10^{-2}$
& $3.45 \times10^{-3}$
& $5.07 \times10^{-5}$ \\
\bottomrule
\end{tabular}
\end{table}

%The results should be interpreted as associational rather than causal. County-level educational attainment, unemployment, demographic composition, and income distributions may be jointly determined by unobserved economic, institutional, and geographic factors. Nevertheless, the analysis demonstrates how the proposed framework can detect both overall and partial associations between scalar predictors and the conditional \Frechet{} mean of a distribution-valued response, while simultaneously assessing the adequacy of a global \Frechet{} regression specification.

\subsection{North Atlantic tropical-cyclone sphere-valued data}
\label{sec:ibtracs}

We apply the proposed procedures to %the North Atlantic subset of
the International Best Track Archive for Climate Stewardship (IBTrACS), which is
publicly available. \footnote{\url{https://www.ncei.noaa.gov/data/international-best-track-archive-for-climate-stewardship-ibtracs/v04r01/access/csv/}}

\paragraph{Data processing}
The raw data consist of storm tracks observed at multiple time points.
To avoid mixing storms from different ocean basins, we restrict the analysis
to storms recorded in the IBTrACS v04r01 North Atlantic file from  1980 to 2025.
%For each storm, the first valid track record is taken as the genesis record.
%Wind speed and central pressure are taken from the U.S. agency fields whenever available and from the corresponding WMO fields otherwise.
We are interested in the lifetime maximum intensity
location of storms, which is defined as the first track point at which the maximum available
sustained wind speed is attained.

Let $\phi_i^\ast$ and $\lambda_i^\ast$ denote the latitude and longitude at
lifetime maximum intensity for the $i$th storm. We represent this location as
a point on the unit sphere, namely $Y_i = \left(\cos(\phi_i^\ast)\cos(\lambda_i^\ast),\cos(\phi_i^\ast)\sin(\lambda_i^\ast),\sin(\phi_i^\ast)  \right)^\top \in \mathbb S^2$.
% \[
%     Y_i
%     =
%     \begin{pmatrix}
%         \cos(\phi_i^\ast)\cos(\lambda_i^\ast)\\
%         \cos(\phi_i^\ast)\sin(\lambda_i^\ast)\\
%         \sin(\phi_i^\ast)
%     \end{pmatrix}
%     \in \mathbb S^2,
% \]
Let $\bm X_i=\left(Z_i,\bm W_i^\top\right)^\top$,
$ Z_i
    =
    \log(1+\mathrm{distance}_i),$
and
$ \bm W_i
    =
    \left(
        \mathrm{wind}_i,\,
        \mathrm{pressure}_i,\,
        \mathrm{translation}_i
    \right)^\top$,
    where $\mathrm{distance}$ denotes the genesis distance to land (measured in kilometers), and
$\bm W_i$ comprises the genesis wind speed (measured in knots), genesis central pressure (measured in
millibars), and genesis
translation speed (measured in knots).
 After removing observations with
missing values in the response or any of the  covariates used below,
the final sample contains $n=685$ storms.

\paragraph{Implementation and results}
To answer whether initially stronger cyclones are more likely to reach peak intensity in certain oceanic regions, or whether faster-moving cyclones tend to peak farther east or north, we first examine whether genesis wind speed, central pressure, and translation speed contain additional information about the \Frechet{} mean of the maximum-intensity location conditional on the genesis distance to land. We conduct the proposed partial significance test, where the unrestricted and reduced regression functions are estimated by local \Frechet{} regression with the same adaptive Epanechnikov product kernel  as in Section \ref{sec:income_data}.

% To reduce the sensitivity to a particular sample partition, we report one
% 50--50 sample-splitting implementation, one five-fold cross-fitted
% implementation, and a Cauchy combination of $J=20$ independently repeated
% five-fold cross-fitting implementations.
As shown in
Table~\ref{tab:ibtracs_results}, the partial significance test does not reject
the null hypothesis. Thus, in this analysis, after conditioning on genesis distance to land, there is no significant
evidence that genesis wind speed, central pressure, and translation speed
jointly provide additional information about the conditional \Frechet{} mean of the
maximum-intensity location.

To proceed, we next focus on the reduced model with the
predictor
\(
    Z_i=\log(1+\mathrm{distance}_i).
\)
  The corresponding
{$p$-values for the global significance test of $Z$ reported in Table \ref{tab:ibtracs_results} show} that genesis
distance to land is strongly associated with the location at which a storm
subsequently reaches its lifetime maximum intensity.

\begin{figure}[h!]
    \centering
    \includegraphics[width=\textwidth]    {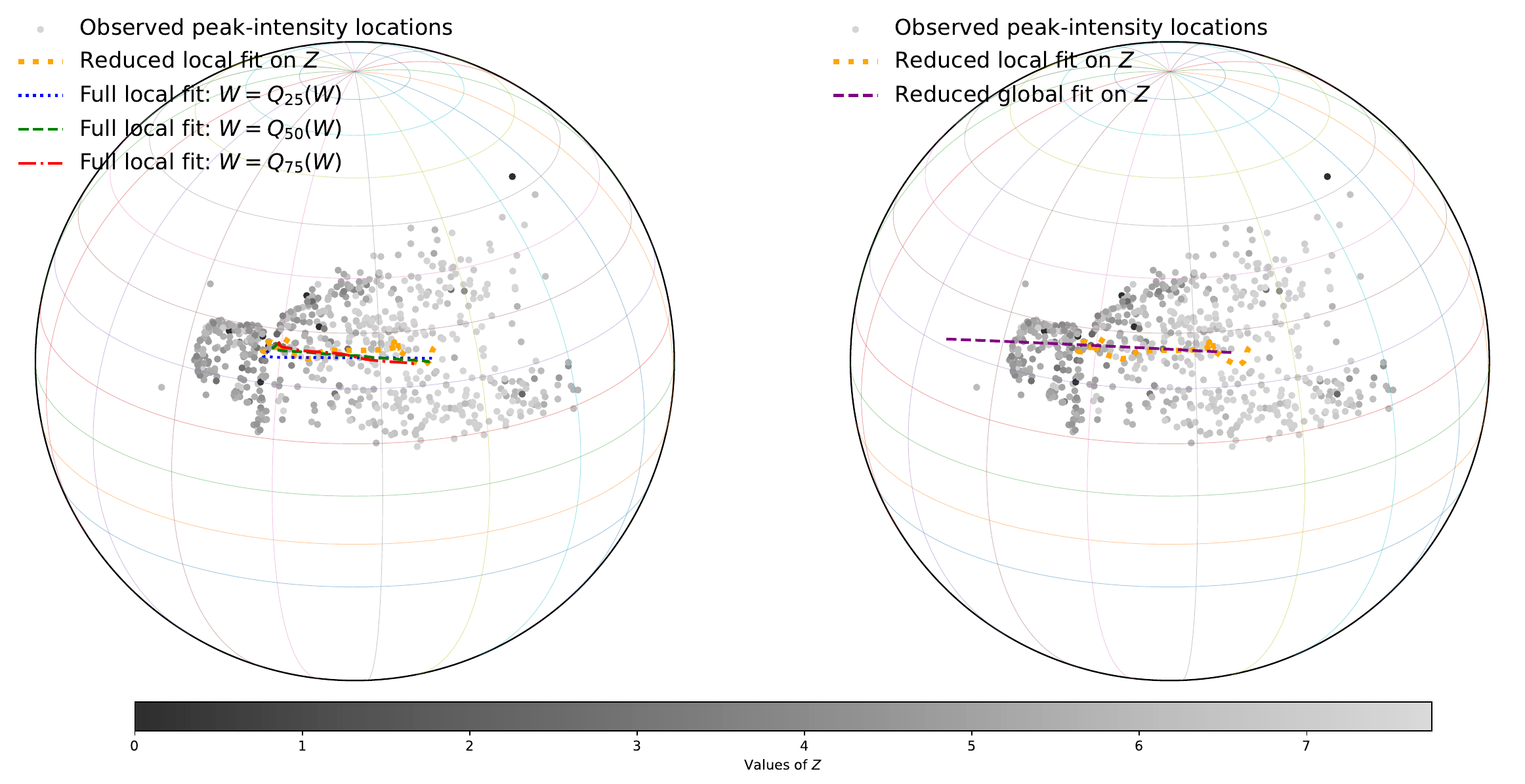}
    \caption{Observed North Atlantic tropical-cyclone locations at lifetime maximum intensity and fitted \Frechet{} regression curves. The left panel compares the reduced local fit in \(Z\) with the full local fits evaluated at the 25th, 50th, and 75th marginal percentiles of \(\bm W\), while the right panel compares the reduced local and global fits in \(Z\).
    %Observed North Atlantic tropical-cyclone locations at lifetime maximum intensity and fitted \Frechet{} regression curves. % as functions of genesis distance to land. Point colors represent $\log(1+\mathrm{distance})$. {The solid curve shows the local \Frechet{} regression fit, and the dashed curve shows the global \Frechet{} regression fit.}
    }
    \label{fig:ibtracs_fit}
\end{figure}

We finally test whether this relationship can be represented by global
\Frechet{} regression.
The results in Table \ref{tab:ibtracs_results} provide evidence against the global \Frechet{} regression specification for the conditional mean as a function of genesis distance to land.
To illustrate the nonlinear effect, we plot  the observed maximum-intensity locations together with the fitted local and global \Frechet{} regression curves in Figure~\ref{fig:ibtracs_fit}.  The local fit moves eastward as genesis occurs farther from land and exhibits noticeable curvature. In contrast, near the lower boundary of $Z$, the global fit
moves outside the region occupied by the observed locations, suggesting
substantial boundary extrapolation. This visual discrepancy is consistent
with the rejection of the global \Frechet{} regression specification,
whereas the local fit remains more closely aligned with the observed spatial  distribution.

\begin{table}[h!]
\centering
\caption{The $p$-values of the proposed multiplier-assisted tests for the North Atlantic IBTrACS data.}
\label{tab:ibtracs_results}
\begin{tabular}{lccc}
\toprule
Testing problem
& Single split
& Cross-fitted
& Cauchy-aggregated \\
\midrule
Partial significance of $\bm W$ given $Z$
& $0.477 $
& $0.616 $
& $0.378 $ \\
Global significance of $Z$
& $1.19\times 10^{-6}$
& $<1.00\times 10^{-6}$
& $<1.00\times 10^{-6}$ \\
 Global specification in $Z$
& $4.78\times 10^{-2}$
& $1.88\times 10^{-2}$
& $1.98\times10^{-4}$ \\
\bottomrule
\end{tabular}
\end{table}

\section{Concluding remarks}\label{sec:conclusion}

We have proposed MATCH, a family of multiplier-assisted tests for conditional hypotheses in \Frechet{} regression with non-Euclidean data objects. MATCH compares an unrestricted conditional \Frechet{} mean with a restricted target and covers global significance, partial significance, and adequacy of the global \Frechet{} regression specification. Independent multipliers applied to held-out losses remove the first-order degeneracy of ordinary loss-gap statistics under the null, while avoiding residual vectors and tangent-space coordinates. Under the regularity conditions stated in the paper, we prove that MATCH procedures are asymptotically valid under the null and consistent against fixed alternatives. Cross-fitting yields a local asymptotic power gain relative to a single split, and repeated cross-fitting with \(p\)-value aggregation reduces sensitivity to a single random partition in implementation.

Several practical considerations should be kept in view when applying the framework. In problems with high-dimensional predictors or complex metric-space-valued responses, performance may depend on the choice of distance, preprocessing, unrestricted and restricted learners, tuning rules, and sample-splitting strategy. These choices can also affect computation, especially when \Frechet{} means or fitted conditional targets are expensive to evaluate repeatedly. In addition, loss-based tests are designed to assess predictive significance or specification adequacy relative to the variables, distances, and models under consideration, and the empirical analyses should therefore be interpreted as associational rather than causal. Future work could develop more automatic learner and tuning selection, scalable implementations for large or computationally intensive metric spaces, diagnostic tools for interpreting departures from restricted specifications, and extensions that combine the proposed testing ideas with study designs or assumptions supporting causal questions.

\clearpage
\appendix
\setcounter{section}{0}
\setcounter{table}{0}
\setcounter{figure}{0}
\renewcommand{\thesection}{S.\arabic{section}}
\renewcommand{\thetable}{S.\arabic{table}}
\renewcommand{\thefigure}{S.\arabic{figure}}
\section{Additional simulation studies}
\label{sec:additional-sim}

\subsection{Comparison of different \(p\)-value merging rules}\label{SEC:p_merging}

\paragraph{Cauchy combination.}
The p-value of the Cauchy combination test of \citet{liu2020cauchy} is defined by
\[
    p_{\mathrm{cauchy}}
    =
    \frac12-\frac1\pi\arctan\left(\frac1J
    \sum_{j=1}^J
    \tan\left\{\pi\left(\frac12-p_j\right)\right\}\right).
\]
This method is attractive %when the \(p\)-values are dependent and
when the
alternative may be sparse across repetitions.

\paragraph{Minimum \(p\)-value merging.} Let
\(
    p_{(1)}\leq p_{(2)}\leq \cdots \leq p_{(J)}
\)
be the ordered \(p\)-values.
The Bonferroni-type minimum \(p\)-value merging rule is
\[
    p_{\min}
    =
    \min\{1,Jp_{(1)}\}.
\]
This rule is valid under arbitrary dependence and is sensitive to the case where
one repetition yields a very small \(p\)-value.

\paragraph{Median \(p\)-value merging.}
The median order-statistic merging rule is
\[
    p_{\mathrm{median}}
    =
    \min\left\{
        1,
        \frac{J}{\lceil J/2\rceil}p_{(\lceil J/2\rceil)}
    \right\}.
\]
Compared with \(p_{\min}\), this rule is less sensitive to a single favorable
split and requires a non-negligible fraction of repetitions to show evidence
against the null.

\paragraph{Arithmetic mean merging.}
Following the averaging rule of \citet{vovk2020combining}, we use
\[
    p_{\mathrm{mean}}
    =
    \min\left\{
        1,
         \frac2J\sum_{j=1}^J p_j
    \right\}.
\]
The factor \(2\) gives a valid merging rule under arbitrary dependence.

\paragraph{Geometric mean merging.}
The geometric mean merging rule is
\[
    p_{\text{geometric}}
    =
    \min\left\{
        1,
        e\left(\prod_{j=1}^J p_j\right)^{1/J}
    \right\}.
\]
The multiplicative factor \(e\) is the dependence-robust calibration for the
geometric mean; see \citet{vovk2020combining}.

\paragraph{Harmonic mean merging.}
The harmonic mean merging rule   is
\[
    p_{\mathrm{harmonic}}
    =
    \min\left\{
        1,
        c_J \left(
        \frac1J\sum_{j=1}^J \frac1{p_j}
    \right)^{-1}
    \right\},
\]
where, for \(J=20\), we take
$ c_J=1.828861\log J$ as recommended by \citet{vovk2020combining}.

\paragraph{\(e\)-value based merging.}
We also consider an \(e\)-value based merging rule. Following
\citet{vovk2021evalues}, a \(p\)-to-\(e\) calibrator is a decreasing function
\(F:[0,1]\to[0,\infty]\) such that \(F(p)\) is an \(e\)-value whenever \(p\) is a
valid \(p\)-value. We use the integrated calibrator
\[
    F(p)
    =
    \int_0^1 \kappa p^{\kappa-1}\,d\kappa
    =
    \frac{1-p+p\log p}{p(-\log p)^2},
    \qquad 0<p<1,
\]
with the convention \(F(1)=1/2\). Define $e_j=F(p_j)$ for $j=1,\ldots,J$.
Since arithmetic averages of \(e\)-values remain valid \(e\)-values under
arbitrary dependence, we set
\(
    \bar e
    =
    J^{-1}\sum_{j=1}^J e_j .
\)
The resulting merged \(p\)-value is
\[
    p_{e}
    =
    \min\left\{
        1,
         1/{\bar e}
    \right\}.
\]

\begin{figure}[h!]
    \centering
    \includegraphics[width=0.49\linewidth]{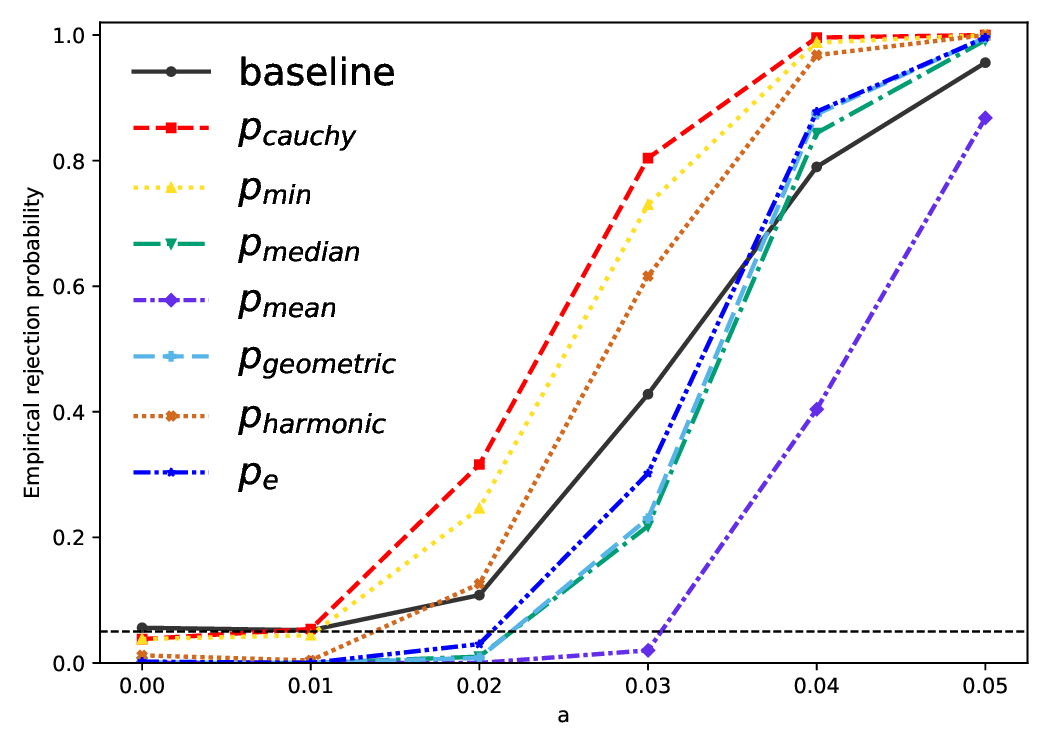}
     \includegraphics[width=0.49\linewidth]{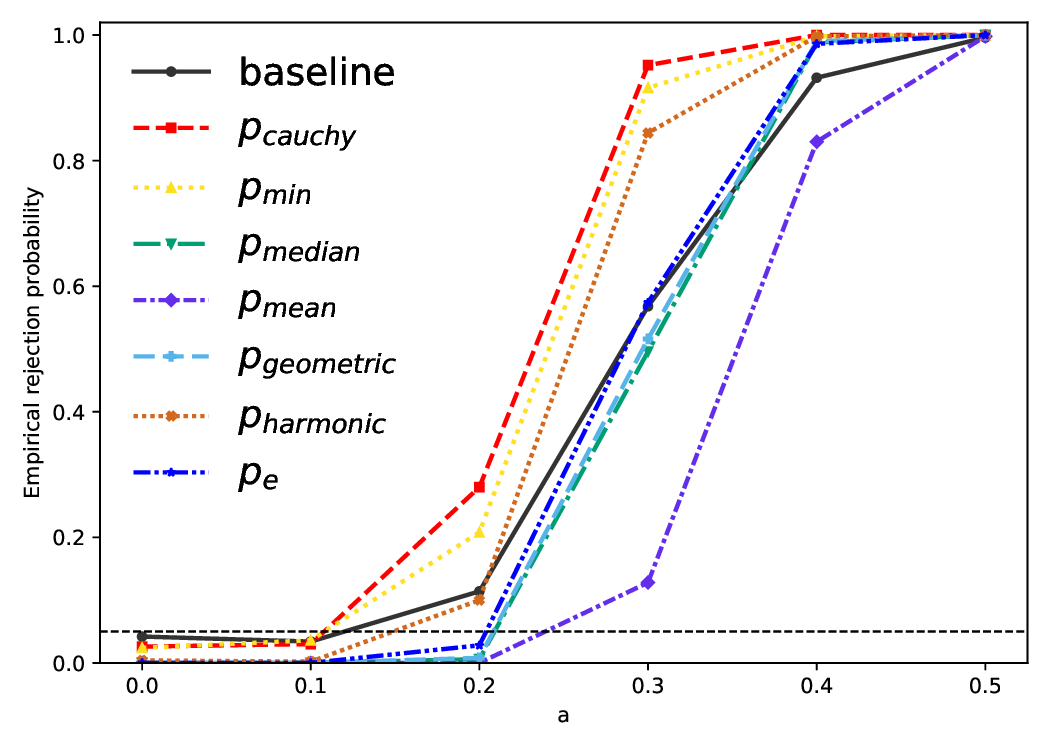}
      \includegraphics[width=0.49\linewidth]{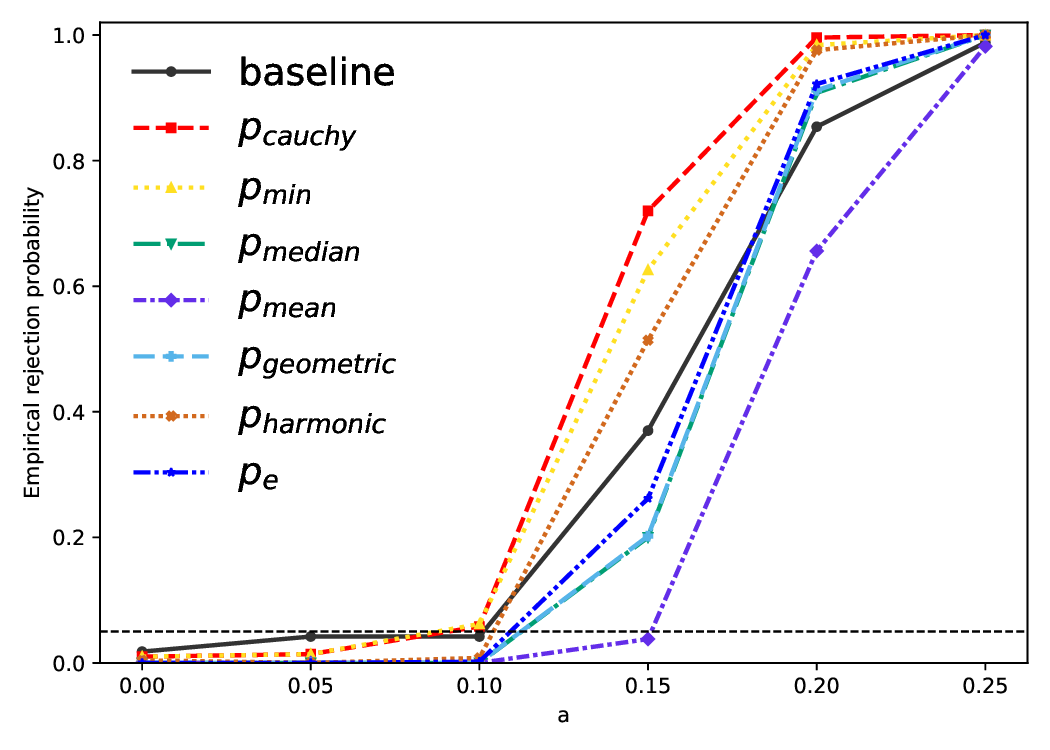}
    \caption{Comparison of different \(p\)-value merging rules under $n=200$ and $J=20$.  The top-left, top-right, and bottom panels   show empirical power curves of
the global, partial, and linearity tests for distributional responses, respectively.}
    \label{fig:pvalue_merging}
\end{figure}

Figure~\ref{fig:pvalue_merging} presents the empirical power curves for the different $p$-value merging methods. Here, the baseline corresponds to the $K$-fold cross-fitted test with $K=5$. Overall, the Cauchy combination achieves the highest power across the considered settings, followed by the minimum $p$-value merging rule and then the harmonic mean merging rule.  The empirical sizes of the merging approaches are close to the nominal level in these settings.

\subsection{Choice of bandwidth in local Fréchet regression}
We further conduct several sensitivity analyses to assess the sensitivity of the proposed tests to implementation choices. The
data-generating mechanisms, sample sizes, signal grids, and local Fréchet
regression estimators are the same as those in the preceding simulation studies,
unless otherwise stated.

We first study the sensitivity to the bandwidth choice.  Let
\(
    h_0=n_{\mathrm{train}}^{-1/(4+q)}
\)
denote the default bandwidth used in the main simulations, where
\(n_{\mathrm{train}}\) is the training sample size and \(q\) is the dimension of
the covariate used in the corresponding local Fréchet estimator.  We compare
two modified bandwidths,
$ h=0.8h_0$ and $ h=1.2h_0$.
The former corresponds to a more localized estimator, while the latter produces
a smoother estimator.  All other components of the test statistic are kept
unchanged. This comparison assesses whether the power curves are sensitive to the   {choice of bandwidth used} in the local \Frechet{} regression.

\begin{figure}[h!]
    \centering
    \includegraphics[width=0.32\linewidth]{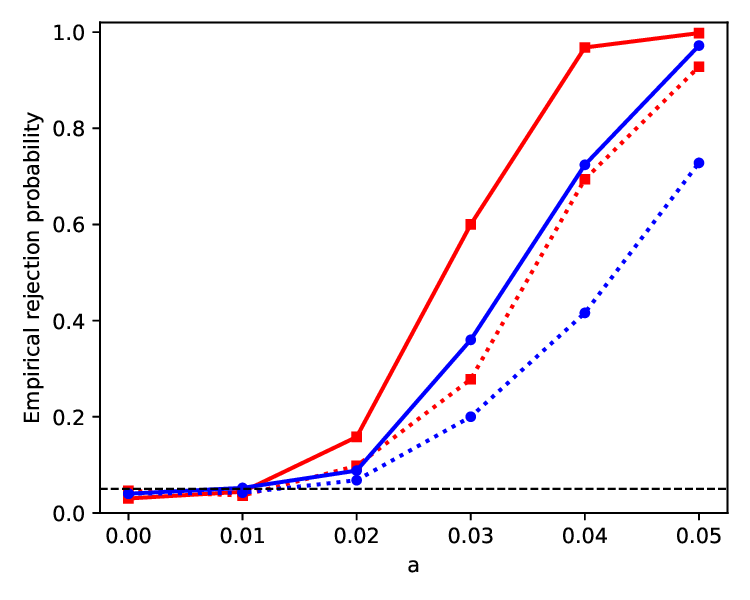}
    \includegraphics[width=0.32\linewidth]{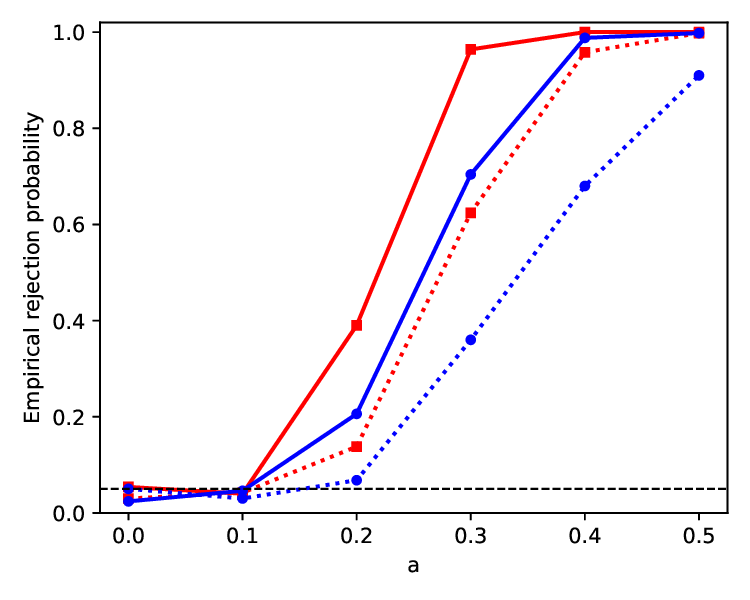}
    \includegraphics[width=0.32\linewidth]{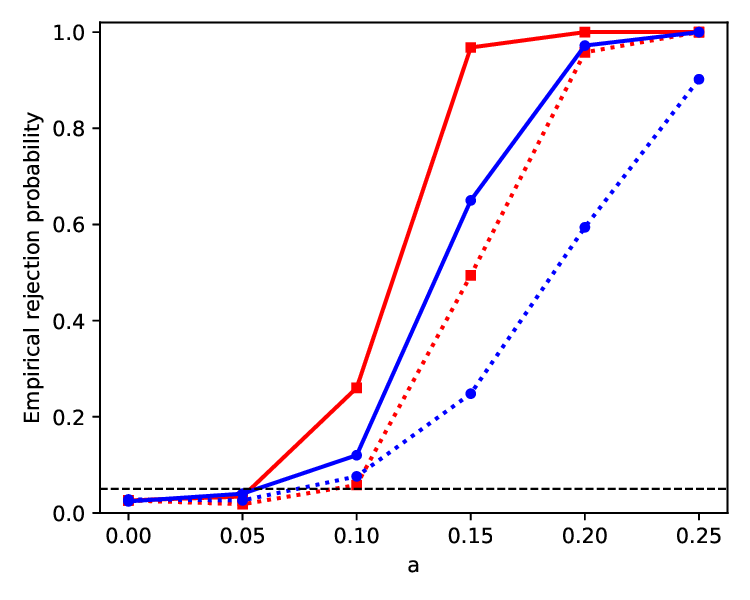}
    \caption{Empirical power curves for distributional responses with bandwidth $h=0.8n_{\mathrm{train}}^{-1/(4+q)}$. Panels from left to right show
the global, partial, and linearity tests. Dotted and solid lines correspond to
\(n=200\) and \(n=400\), respectively. Blue circles denote \(\widetilde T_n\), and red
squares denote \(\widetilde T_{n,K}\). The dashed horizontal line marks \(\alpha=0.05\).}
    \label{fig:bandwidth0.8}
\end{figure}

\begin{figure}[h!]
    \centering
    \includegraphics[width=0.32\linewidth]{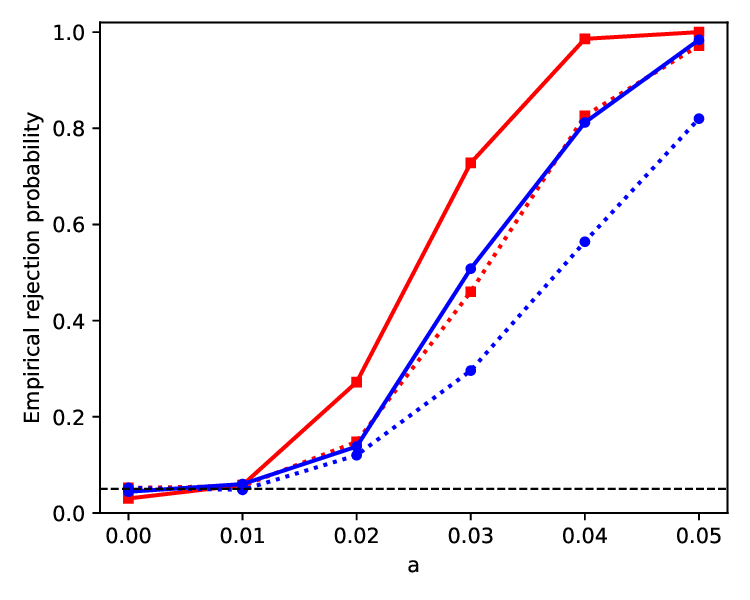}
    \includegraphics[width=0.32\linewidth]{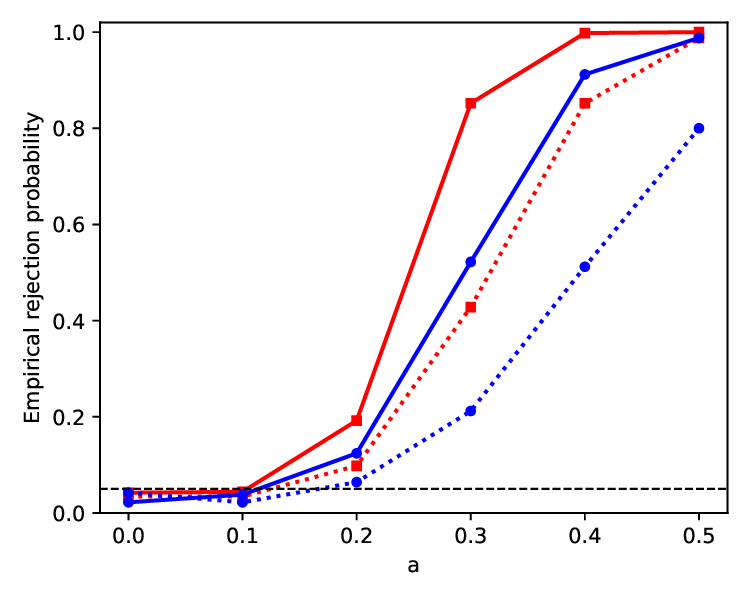}
    \includegraphics[width=0.32\linewidth]{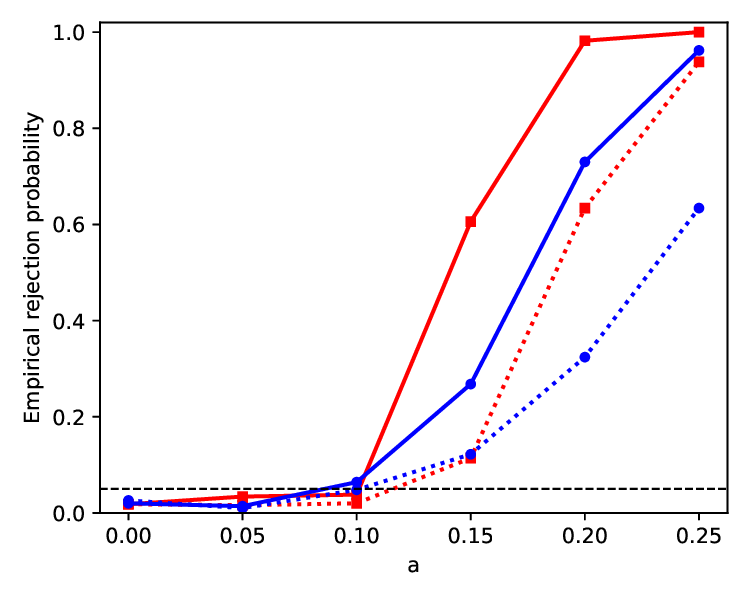}
    \caption{Empirical power curves for distributional responses with bandwidth $h=1.2n_{\mathrm{train}}^{-1/(4+q)}$. Panels from left to right show
the global, partial, and linearity tests. Dotted and solid lines correspond to
\(n=200\) and \(n=400\), respectively. Blue circles denote \(\widetilde T_n\), and red
squares denote \(\widetilde T_{n,K}\). The dashed horizontal line marks \(\alpha=0.05\).}
    \label{fig:bandwidth1.2}
\end{figure}

Figures~\ref{fig:bandwidth0.8}-\ref{fig:bandwidth1.2} show that both the empirical size and power remain broadly stable over a moderate range of bandwidth choices. Intuitively, when the bandwidth is excessively large, the local estimator is oversmoothed, leading to lower variance but greater bias. As a result, deviations from the null model may be attenuated, reducing the power of the test. In contrast, when the bandwidth is too small, the estimator has lower bias but substantially higher variance. The resulting instability in the estimated nuisance functions increases the variability of the held-out loss differences and weakens the signal-to-noise ratio, which  also leads to lower power.

\subsection{Choice of $K$ in cross-fitting}
We examine the effect of the number of folds in the cross-fitting
procedure.  In the main simulations we use \(K=5\).  Here we additionally
consider $K=2$ and $K=10$.
When $K=2$, the training and evaluation samples are of approximately equal size within each fold, while $K=10$ uses larger training samples and smaller evaluation folds.
%The case \(K=2\) uses larger evaluation folds but smaller training samples, whereas \(K=10\) uses larger training samples but smaller evaluation folds.  %This comparison assesses whether the power curves are sensitive to the particular choice of \(K\).

\begin{figure}[h!]
    \centering
    \includegraphics[width=0.32\linewidth]{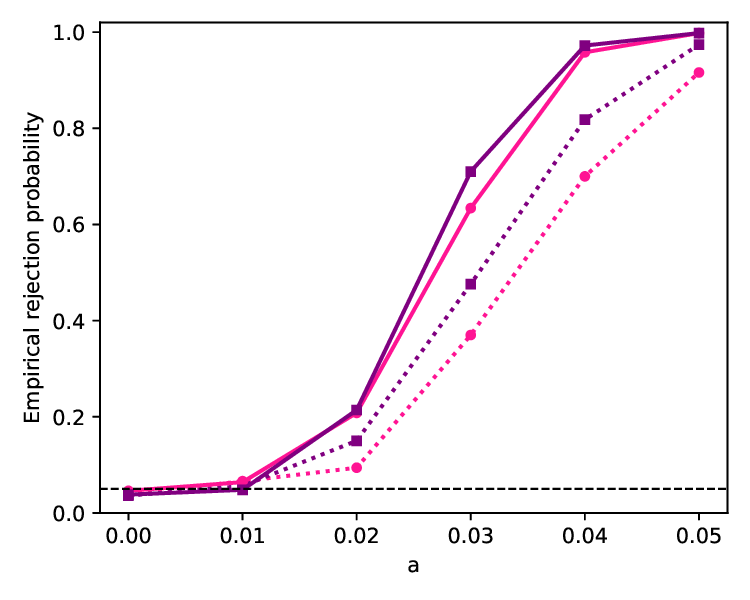}
    \includegraphics[width=0.32\linewidth]{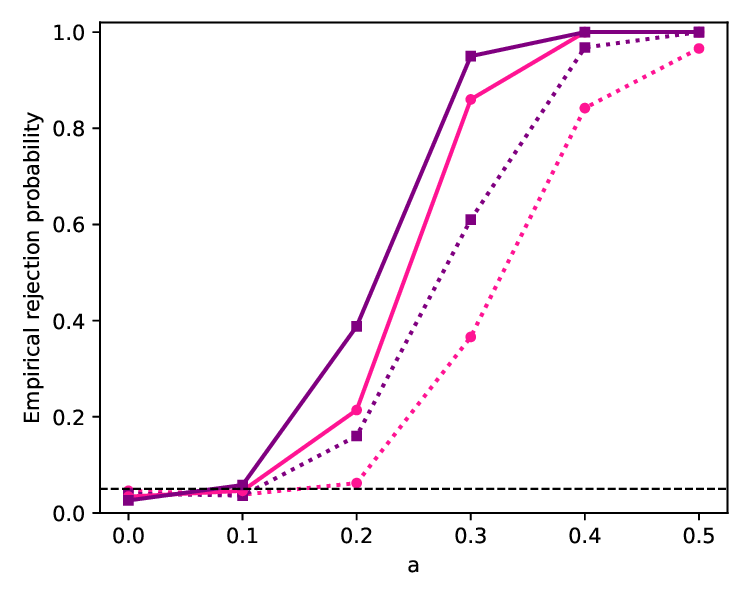}
    \includegraphics[width=0.32\linewidth]{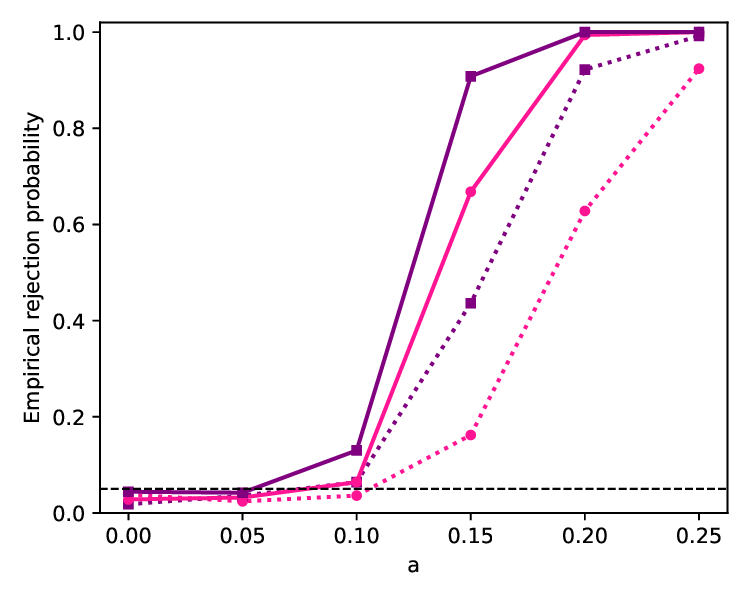}
    \caption{Empirical power curves for distributional responses. Panels from left to right show
the global, partial, and linearity tests. Dotted and solid lines correspond to
\(n=200\) and \(n=400\), respectively. Pink circles denote  \(\widetilde T_{n,K}\) with $K=2$, and purple squares denote $\widetilde T_{n,K}$ with $K=10$. The dashed horizontal line marks \(\alpha=0.05\).}
    \label{fig:K}
\end{figure}

Figure~\ref{fig:K} shows that the test with $K=2$ has slightly lower power, likely because the smaller training samples lead to larger nuisance estimation errors. By contrast, the performance with $K=10$ is very similar to that with $K=5$. Balancing statistical power and computational efficiency, we  recommend $K=5$ as the default choice.

\subsection{Choice of multipliers}
We replace the Gaussian multipliers  {with} Gamma multipliers.  For the
 statistic \(\widetilde T_n\), we generate independent multipliers
\[
    \varphi_i,\psi_i
    \overset{\mathrm{i.i.d.}}{\sim}
    \operatorname{Gamma}(2,1/2),
\]
satisfying $E\varphi_i=E\psi_i=1$, $\operatorname{Var}(\varphi_i)=\operatorname{Var}(\psi_i)=1/2$,
% \[
%     E\varphi_i=E\psi_i=1,
%     \qquad
%     \operatorname{Var}(\varphi_i)=\operatorname{Var}(\psi_i)=1/2,
% \]
matching the moment conditions  in the manuscript.  The corresponding
cross-fitted statistic is obtained by using the same Gamma multipliers in
\(\widetilde T_{n,K}\).
\begin{figure}[h!]
    \centering
    \includegraphics[width=0.32\linewidth]{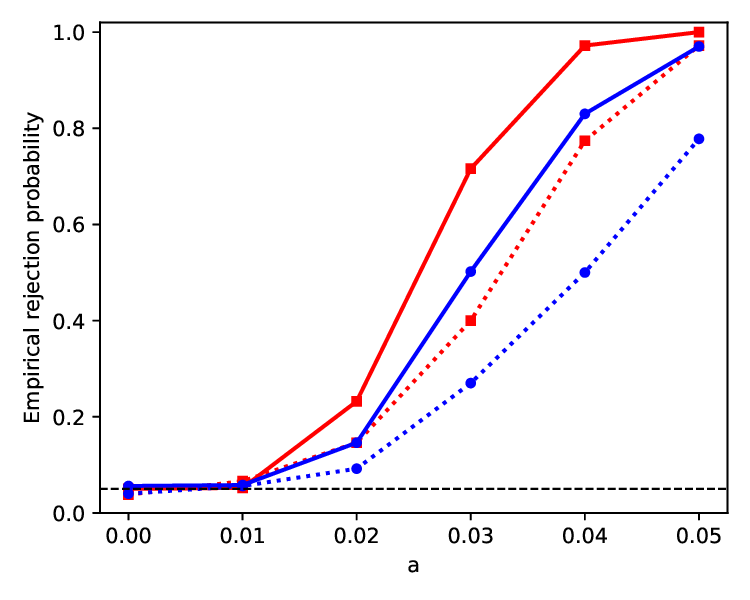}
    \includegraphics[width=0.32\linewidth]{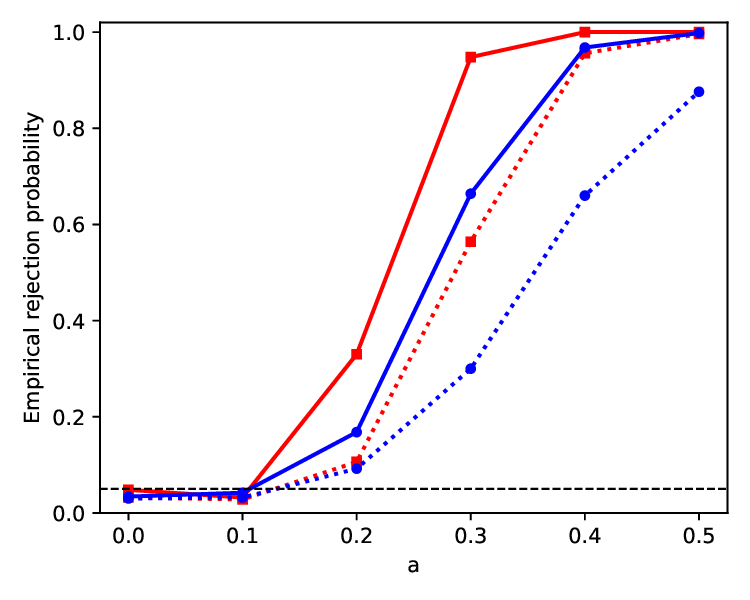}
    \includegraphics[width=0.32\linewidth]{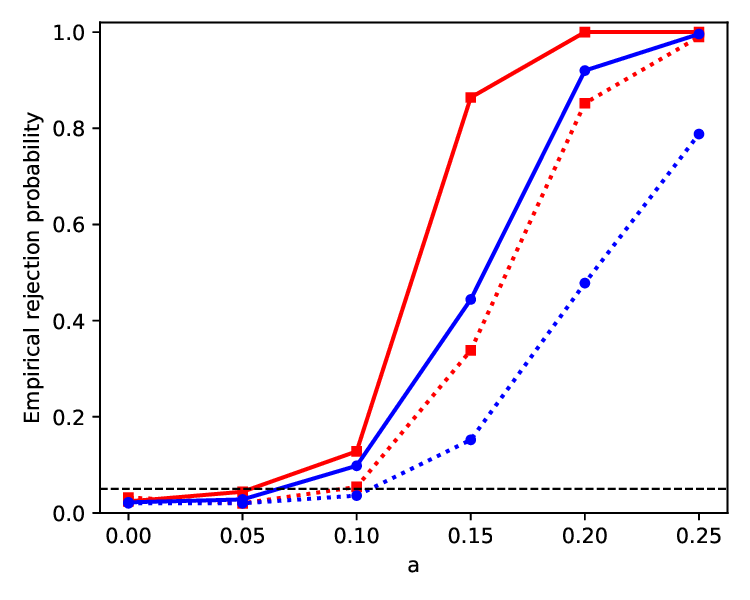}
    \caption{Empirical power curves for distributional responses. Panels from left to right show
the global, partial, and linearity tests. Dotted and solid lines correspond to
\(n=200\) and \(n=400\), respectively. Blue circles denote  \( \widetilde T_{n}\) with gamma multipliers, and red squares denote $\widetilde T_{n,K}$ with gamma multipliers. The dashed horizontal line marks \(\alpha=0.05\).}
    \label{fig:gamma_multiplier}
\end{figure}

Figure~\ref{fig:gamma_multiplier} shows that the empirical power curves obtained with Gamma multipliers are very similar to those obtained with Gaussian multipliers, suggesting limited sensitivity to this multiplier choice in the simulated settings.

\subsection{Alternative approach: asymmetric randomization}
We also consider the alternative  multiplier statistic $T_n^\dagger$ discussed in Remark \ref{remark2}.
% \[
%     T_n^\dagger
%     =
%     \frac1{n_2}
%     \sum_{i\in\mathcal D_2}
%     \left[
%         \varphi_i d^2\{Y_i,\widehat m(\bm X_i)\}
%         -
%         d^2\{Y_i,\widehat g(\bm X_i)\}
%     \right],
% \]
% where the multiplier satisfies \(E\varphi_i=1\) and
% \(\operatorname{Var}(\varphi_i)=1\).
The cross-fitted counterpart is defined as
\[
    T_{n,K}^\dagger
    =
    \frac1n
    \sum_{k=1}^K
    \sum_{i\in\mathcal I_k}
    \left[
        \varphi_i' d^2\{Y_i,\widehat m_{-k}(\bm X_i)\}
        -
        d^2\{Y_i,\widehat g_{-k}(\bm X_i)\}
    \right]-\max\{b_{n,K},0\}.
\]

\begin{figure}[h!]
    \centering
    \includegraphics[width=0.32\linewidth]{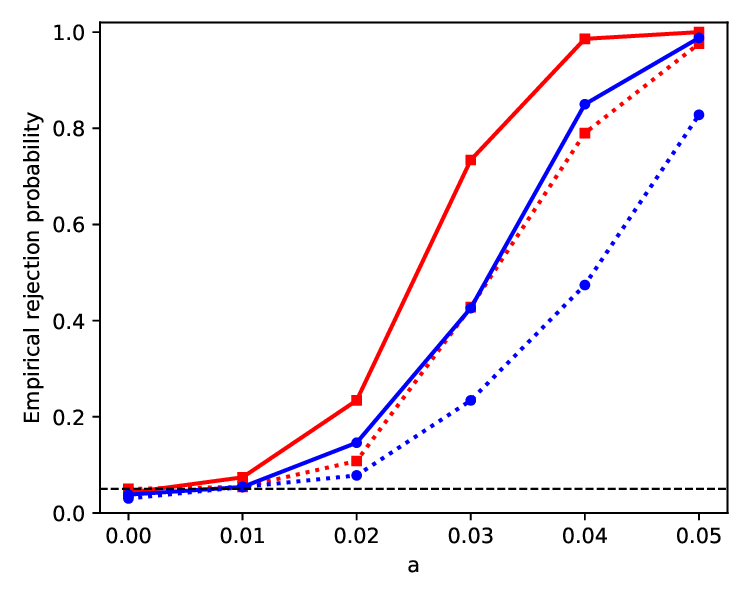}
    \includegraphics[width=0.32\linewidth]{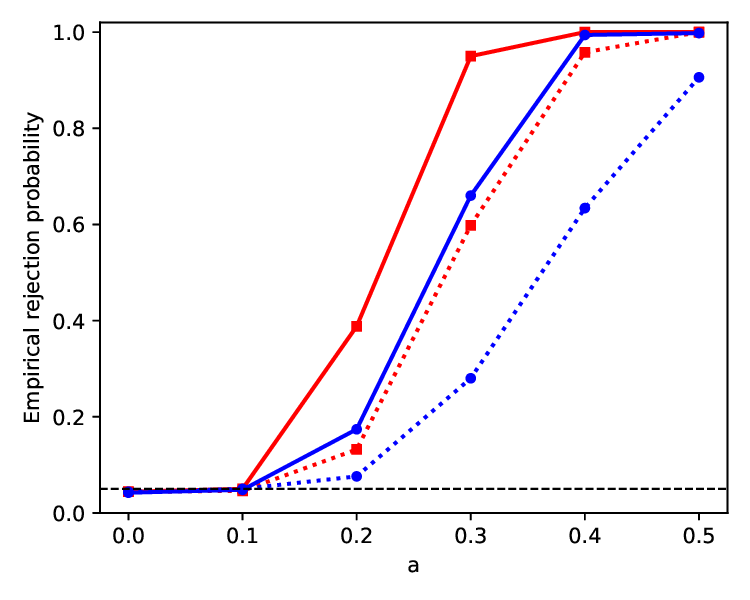}
    \includegraphics[width=0.32\linewidth]{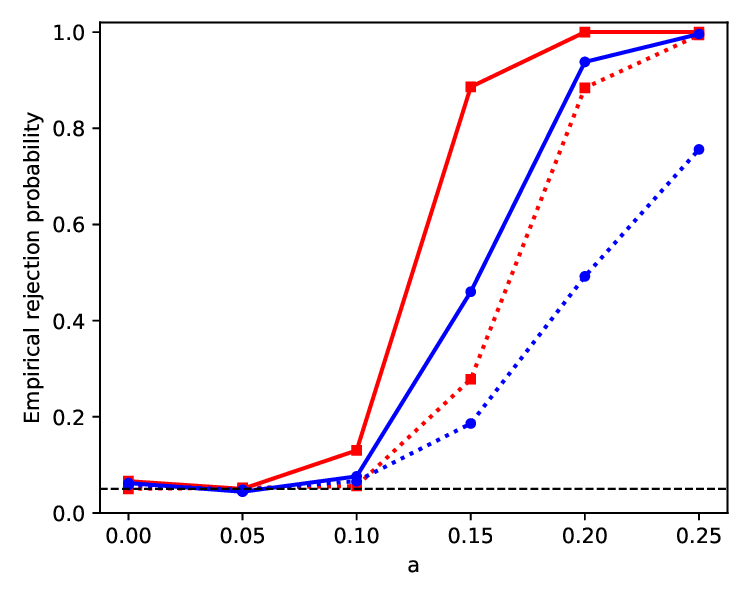}
    \caption{Empirical power curves for distributional responses. Panels from left to right show
the global, partial, and linearity tests. Dotted and solid lines correspond to
\(n=200\) and \(n=400\), respectively. Blue circles denote  \(T_{n}^{\dagger}\), and red squares denote $T_{n,K}^\dagger$. The dashed horizontal line marks \(\alpha=0.05\).}
    \label{fig:asymmetric_multiplier}
\end{figure}

Figure~\ref{fig:asymmetric_multiplier} shows that the alternative  statistics   {perform similarly to} the original statistics, consistent with the discussion in Remark \ref{remark2}.

\section{Discussion of Assumption (A1)}\label{sec:A1}

\subsection{Euclidean responses}

Let \(\Om=\R^q\), or assume that all conditional, marginal, partial conditional, and signed weighted means considered below lie in a closed convex set \(\Om\subseteq\R^q\). Let
$ d(\omega_1,\omega_2)=\|\omega_1-\omega_2\|_2 $.
Since $m(x)=\E(Y\mid X=x)$,
 for any \(\omega\in\Om\),
\begin{align*}
    M(\omega,x)-M\{m(x),x\}
    &=
    \E\{\|Y-\omega\|^2-\|Y-m(x)\|^2\mid X=x\}  \\
    &=
    \|\omega-m(x)\|^2
    -2\left\langle \omega-m(x),\E\{Y-m(x)\mid X=x\}\right\rangle \\
    &=
    \|\omega-m(x)\|^2 .
\end{align*}
Therefore the local quadratic upper bound for \( {M}(\omega,x)\) holds globally with \(C=1\).
 {A similar calculation verifies} the bound for the common three choices of \(G\).

\begin{itemize}[leftmargin=2em]
\item[(i)] \textbf{Global significance.}
$ G(\omega,x)=\E\|Y-\omega\|^2$,
  \(g=\E Y\), thus
\[
    G(\omega,x)-G(g,x)=\|\omega-g\|^2 .
\]

\item[(ii)] \textbf{Partial significance.}
$G(\omega,x)=\E\{\|Y-\omega\|^2\mid Z=z\}$,
  \(g(x)=f(z)=\E(Y\mid Z=z)\), thus
\[
    G(\omega,x)-G(g(x),x)
    =
    \|\omega-g(x)\|^2 .
\]

\item[(iii)] \textbf{Global linearity.}
$G(\omega,x)=\E\{s(X,x)\|Y-\omega\|^2\}$,  \(\E\{s(X,x)\}=1\),
assume that \(G(\cdot,x)\) has a unique minimizer in \(\Om\) and that the signed weighted mean \(\E\{s(X,x)Y\}\) belongs to \(\Om\). Then \(g(x)=\E\{s(X,x)Y\}\),
and a direct expansion gives
\[
    G(\omega,x)-G(g(x),x)
    =
    \|\omega-g(x)\|^2 .
\]
\end{itemize}
Hence, the Euclidean case satisfies Assumption (A1) exactly with \(C=1\). %If the main theorem also assumes finite diameter of \(\Om\), one may additionally restrict \(\Om\) to a compact Euclidean set. This compactness is not needed for the local quadratic bound itself.

\subsection{Spherical responses with geodesic distance}

Let \(\Om=S^{r-1}\) and let \(d=d_g\) be the spherical geodesic distance. For a generic objective \(H\in\{M,G\}\), write its target as \(h(x)\), where \(h(\cdot)=m(\cdot)\) for \(H(\omega,\bm x)=M(\omega,\bm x)\) and \(h(\cdot)=g(\cdot)\) for \(H(\omega,\bm x)=G(\omega,\bm x)\). The squared geodesic loss
$\omega\mapsto d_g^2(Y,\omega)$
is smooth away from the cut locus of \(Y\). On the sphere, the cut locus of \(\omega\) is its antipodal point \(-\omega\). A convenient sufficient condition is to keep the local neighborhood of \(h(x)\) uniformly away from antipodal singularities, together with the stated bounded-weight and uniqueness conditions.

One way to formalize the cut-locus condition is: there exist \(\eta>0\) and \(\epsilon>0\) such that,
\begin{align}\label{EQ:sufficient0}
    \P_H\left(
        \inf_{\omega\in B_g(h(\bm x),\eta)}
        d_g(Y,-\omega)>\epsilon
    \right)=1.
\end{align}
Here, $B_g(h(\bm x),\eta)$ denotes the geodesic ball centered at $h(\bm x)$,
and $-\omega$ is the antipodal point of $\omega$ on the sphere.  The probability law $\P_H$ in the above condition depends on the
objective function under consideration. More explicitly, for the full conditional \Frechet{} objective
\(
    M(\omega,\bm x)=\E\{d_g^2(Y,\omega)\mid \bm X=\bm x\},
\)
the relevant probability law is the conditional law of \(Y\) given \(\bm X=\bm x\).
Thus, the condition is
\[
    \P\left(
        \inf_{\omega\in B_g(m(\bm x),\eta)}
        d_g(Y,-\omega)>\epsilon
        \,\middle|\, \bm X=\bm x
    \right)=1.%,\quad {\text{uniformly in \(\bm x\)}}.
\]
For the global significance test, the restricted target is the global \Frechet{}
mean
\(
    g(\bm x)\equiv \omega_0
    =
    \arg\min_{\omega\in\Omega}\E d_g^2(Y,\omega).
\)
The relevant probability law is the marginal law of \(Y\). Hence the condition is
\[
    \P\left(
        \inf_{\omega\in B_g(\omega_0,\eta)}
        d_g(Y,-\omega)>\epsilon
    \right)=1 .
\]
For the partial significance test, write \(\bm X=(\bm Z^\top,\bm W^\top)^\top\)
and \(\bm x=(\bm z^\top,\bm w^\top)^\top\). The restricted target is
\(
    g(\bm x)=f(\bm z)
    =
    \arg\min_{\omega\in\Omega}
    \E\{d_g^2(Y,\omega)\mid Z=\bm z\}.
\)
The relevant probability law is the conditional law of \(Y\) given \(\bm Z=\bm z\).
Thus, the condition is
\[
    \P\left(
        \inf_{\omega\in B_g(f(\bm z),\eta)}
        d_g(Y,-\omega)>\epsilon
        \,\middle|\, \bm Z=\bm z
    \right)=1.%,\quad {\text{uniformly in \(\bm z\)}}.
\]
For the linearity test, the restricted target is
\(
    g(\bm x)=m_0(\bm x)
    =
    \arg\min_{\omega\in\Omega}
    \E\{s(X,\bm x)d_g^2(Y,\omega)\},
\)
where
\(
    s(\bm X,\bm x)=1+(\bm X-\bm \mu)^\top\mathbf\Sigma^{-1} {(\bm x-\bm \mu)},
\) and $\bm \mu=\E\bm X$.
Here \(s(\bm X,\bm x)\) may be signed, so it does not generally define a probability
law. Therefore,   we impose
the   condition under the original joint law:
\[
    \P\left(
        \inf_{\omega\in B_g(m_0(\bm x),\eta)}
        d_g(Y,-\omega)>\epsilon
    \right)=1.%, \quad {\text{uniformly in \(\bm x\)}}.
\]
 In addition, to control the weighted Hessian, assume
\(
    \sup_{\bm x\in\mathcal X}\E |s(\bm X,\bm x)|<\infty .
\)
% Then the weighted objective also has a uniformly bounded Hessian in the local
% neighborhood of \(m_0(\bm x)\).

This condition
means that, with probability one, $Y$ stays uniformly away from the cut loci of
all points $\omega$ in a small neighborhood of $h(\bm x)$. Consequently, the squared
geodesic loss $\omega\mapsto d_g^2(Y,\omega)$ is smooth on this neighborhood and
has uniformly bounded Hessian.

A stronger but easier-to-check version is
\begin{align}
    \label{EQ:sufficient1}
    \P_H \{Y\in B_g(h(\bm x),R)\}=1
    \quad\text{for some }R<\pi/2.
\end{align}
Under this condition, for all \(\omega\in B_g(h(\bm x),\eta)\), $\eta<\pi-R$,  the distance \(d_g(Y,\omega)\) is bounded away from \(\pi\).
% On this region the Hessian of
% \[
%     \omega\mapsto d_g^2(Y,\omega)
% \]
% is uniformly bounded.

Let \(p=h(\bm x)\) and \(\bm v=\Log_p(\omega)\in T_pS^{r-1}\). Then
\(\|\bm v\|=d_g(\omega,h(\bm x))\). Define the constant-speed geodesic
$\gamma(t)=\Exp_p(t\bm v)$, $t\in[0,1]$,
so that \(\gamma(0)=h(\bm x)\) and \(\gamma(1)=\omega\). Let
\(
    F_{\bm x}(t)=H(\gamma(t),\bm x).
\)
Taylor's formula with integral remainder gives
\[
    H(\omega,\bm x)-H(h(\bm x),\bm x)
    =
    F_{\bm x}(1)-F_{\bm x}(0)
    =
    F_{\bm x}'(0)+\int_0^1(1-t)F_{\bm x}''(t)\,dt .
\]
Since \(h(\bm x)\) is the local minimizer of \(H(\cdot,\bm x)\), the first-order term
vanishes:
\(
    F_{\bm  x}'(0)=0.
\)
Moreover,
\[
    F_{\bm x}''(t)
    =
    \nabla^2 H(\gamma(t),\bm x)[\dot\gamma(t),\dot\gamma(t)].
\]
Under (\ref{EQ:sufficient1}), let $\epsilon_0=\pi-R-\eta>0$, $\epsilon=\min\{\pi/4,\epsilon_0\}$,  for the constant \(C_0=2\max\{1,(\pi-\epsilon)\cot\epsilon\}<\infty\),
\[
    |F_{\bm x}''(t)|
    \leq
    C_0\|\dot\gamma(t)\|^2
    =
    C_0 d_g^2\{\omega,h(\bm x)\}.
\]
Therefore,
\[
    H(\omega,\bm x)-H\{h(\bm x),\bm x\}
    \leq
    \int_0^1(1-t)C_0d_g^2\{\omega,h(\bm x)\}\,dt
    =
    \frac{C_0}{2}d_g^2\{\omega,h(\bm x)\}.
\]

% In geodesic normal coordinates at \(h(\bm x)\), Taylor expansion along the geodesic from \(h(\bm x)\) to \(\omega\) gives
% \[
%     H(\omega,\bm x)-H\{h(\bm x),\bm x\}
%     \le C d_g^2\{\omega,h(\bm x)\},
%     \quad
%     d_g\{\omega,h(\bm x)\}\le \eta,
% \]
% since the first derivative vanishes at the Fr\'echet mean or local minimizer \(h(\bm x)\), and the nonnegativity follows from the local minimality of \(h(\bm x)\). %Thus Assumption (A1) holds on the sphere under uniform cut-locus avoidance.

\subsection{SPD responses}

Let \(\Om=S_{++}^r\), the cone of \(r\times r\) symmetric positive definite matrices.
Consider the Log-Euclidean metric,
\[
    d_{\mathrm{LE}}(\mathbf A,\mathbf B)
    =
    \|\log \mathbf A-\log \mathbf B\|_F .
\]
Here, \(\log \mathbf A\) means the \emph{principal matrix logarithm} of the SPD matrix \(\mathbf A\).%, not the entrywise logarithm.
Since \(\mathbf A\in S_{++}^r\), it has a spectral decomposition
\[
    \mathbf A=\mathbf U\operatorname{diag}(\lambda_1,\ldots,\lambda_r)\mathbf U^\top,
    \quad
    \lambda_j>0.
\]
The matrix logarithm is defined by applying the scalar logarithm to the eigenvalues:
\[
    \log \mathbf A
    =
    \mathbf U\operatorname{diag}(\log\lambda_1,\ldots,\log\lambda_r)\mathbf U^\top .
\]
%This matrix is symmetric.
Conversely, for any symmetric matrix
\[
    \mathbf S=\mathbf U\operatorname{diag}(s_1,\ldots,s_r)\mathbf U^\top,
\]
the matrix exponential
\[
    \exp \mathbf S
    =
    \mathbf U\operatorname{diag}(e^{s_1},\ldots,e^{s_r})\mathbf U^\top
\]
is SPD. Thus
\(
    \log:S_{++}^r\to \Sym(r)
\)
is a one-to-one smooth transformation from the SPD cone to the Euclidean vector space of symmetric matrices, with inverse map \(\exp:\Sym(r)\to S_{++}^r\).

The Log-Euclidean metric is exactly the Euclidean Frobenius distance after this logarithmic transformation.
% In other words, if
% \(
%     \mathbf Z=\log \mathbf Y\),
%     \(\mathbf U_\omega=\log\omega,
% \)
% then
% \(
%     d_{\mathrm{LE}}^2(Y,\omega)
%     =
%     \|\mathbf Z-\mathbf U_\omega\|_F^2 .
% \)
Therefore, the Fr\'echet regression problem under \(d_{\mathrm{LE}}\)  {can be treated as a Euclidean} squared-loss problem in the log-domain \(\Sym(r)\).

Suppose
\[
    \theta(\bm x)=\E(\log \mathbf Y\mid \bm X=\bm x),
    \quad
    m(\bm x)=\exp\{\theta(\bm x)\}.
\]
Then,
\begin{align*}
    M(\omega,\bm x)-M(m(\bm x),\bm x)
    &=
    \E\{\|\log \mathbf Y-\log\omega\|_F^2
      -\|\log \mathbf Y-\theta(\bm x)\|_F^2\mid \bm X=\bm x\}  \\
    &=
     {\|\log\omega-\theta(\bm x)\|_F^2}  \\
    &=
    d_{\mathrm{LE}}^2(\omega,m(\bm x)).
\end{align*}
% Thus, the local quadratic upper bound for \(M(\omega,\bm x)\) holds globally with \(C=1\).
The same exact identity holds for \(G(\omega,\bm x)\) after replacing \(\theta(\bm x)\) by the corresponding marginal, partial conditional, or weighted mean of \(\log \mathbf Y\). Hence, Assumption (A1) holds globally with \(C=1\) for the Log-Euclidean metric.

\subsection{  Wasserstein distributional responses}

Let \(\Om \) be the space of probability
measures with finite second moments on  a compact domain,  and let \(d=\mathcal W_2\) denote the Wasserstein metric. For one-dimensional distributions, the quantile representation gives the isometry
\[
    \mathcal W_2^2(\mu,\nu)
    =
    \|Q_\mu-Q_\nu\|_{L^2[0,1]}^2,
\]
where \(Q_\mu\) is the quantile function of \(\mu\). Thus the problem becomes an ordinary squared-loss problem in the Hilbert space \(L^2[0,1]\), restricted to the closed convex cone of nondecreasing quantile functions.

Define $ q(\bm x)=\E(Q_Y\mid \bm X=\bm x),$
 $Q_{m(\bm x)}=q(\bm x)$.
Then, for any \(\omega\in\Om\),
\begin{align*}
    M(\omega,\bm x)-M\{m(\bm x),\bm x\}
    &=
    \E\{\|Q_Y-Q_\omega\|_{L^2[0,1]}^2
       -\|Q_Y-q(\bm x)\|_{L^2[0,1]}^2\mid \bm X=\bm x\} \\
    &=
    \|Q_\omega- {q(\bm x)}\|_{L^2[0,1]}^2 \\
    &=
    \mathcal W_2^2(\omega,m(\bm x)).
\end{align*}
Therefore the local quadratic upper bound for \(M(\omega,\bm x)\) holds globally with \(C=1\). For \(G(\omega,\bm x)\), the same identity holds with \(q(\bm x)\) replaced by the corresponding marginal, partial conditional, or weighted mean quantile, provided this mean is a valid quantile function. For the global linearity objective with signed weights, we impose this validity as an additional condition; otherwise the exact Hilbert-space identity applies only to the unconstrained \(L^2\) target, while the constrained Wasserstein target requires a projection or uniqueness argument. Under this condition, Assumption (A1) holds with $C=1$ in the one-dimensional Wasserstein case.

%\subsection{Multidimensional Wasserstein distributional responses}

For multidimensional   distributions,
 there is no global quantile
isometry   into a Hilbert space. Therefore,
the local quadratic upper bound is not automatic and requires regularity of the
optimal transport maps around the Fr\'echet target.
For a generic objective \(H\in\{M,G\}\), write $h(\bm x)=\arg\min_{\omega\in\Omega}H(\omega,\bm x)$.
In the following, we provide a sufficient regularity condition for Assumption (A1).

% For the Taylor expansion below, it is enough to assume that along the local
% Wasserstein geodesics connecting \(h(\bm x)\) and \(\omega\), the map
% \[
%     t\mapsto H(\omega_t,\bm x)
% \]
% is twice differentiable and satisfies the uniform bound
% \[
%     \left|
%     \frac{d^2}{dt^2}H(\omega_t,\bm x)
%     \right|
%     \le
%     C W_2^2\{\omega,h(\bm x)\}.
% \]
% This condition follows, for example, from uniform \(\mathcal C^2\) regularity of the
% relevant Brenier maps and Kantorovich potentials.

\begin{comment}
Assume that, uniformly in the covariate value defining \(H\), there exist
\(\eta>0\), a compact convex set \(K\subset\mathbb R^d\), and constants
\(0<c<C<\infty\) such that \(Y\), \(h(\bm x)\), and all
\(\omega\in B_{W_2}\{h(\bm x),\eta\}\) are absolutely continuous distributions
supported on \(K\), with densities bounded as
\[
    c\le \rho \le C
    \quad\text{on }K.
\]
In addition, assume that the relevant Brenier maps and Kantorovich potentials
between distributions in this local class are uniformly \(\mathcal C^2\), with uniformly
bounded derivatives up to order two. For weighted linear Fr\'echet objectives,
where the weight \(s(X,\bm x)\) may be signed, additionally assume
\[
    \sup_{\bm x\in\mathcal X}\mathbb E|s(X,\bm x)|<\infty.
\]
\end{comment}

% Under these assumptions, the map
% \[
%     \omega\mapsto W_2^2(Y,\omega)
% \]
% is twice differentiable along Wasserstein geodesics in a neighborhood of
% \(h(\bm x)\), and its second derivative is uniformly bounded. More precisely,
For any \(\omega\in B_{\mathcal W_2}(h(\bm x),\eta)\), let \(T\) be the optimal transport
map from \(h(\bm x)\) to \(\omega\). Define the constant-speed Wasserstein
pushforward
$\omega_t=\{(1-t)\mathrm{Id}+tT\}_{\#}h(\bm x)$, $t\in[0,1]$.
Then, $\omega_0=h(\bm x)$, $\omega_1=\omega$, and
\(
    \mathcal W_2^2\{\omega,h(\bm x)\}
    =
    \int \|T(u)-u\|^2\,dh(\bm x)(u).
\)
Assume that along the local
Wasserstein geodesics connecting \(h(\bm x)\) and \(\omega\), the map
\(
    t\mapsto H(\omega_t,\bm x)
\)
is twice differentiable and satisfies the uniform bound
\[
    \left|
    \frac{d^2}{dt^2}H(\omega_t,\bm x)
    \right|
    \le
    C_0 W_2^2\{\omega,h(\bm x)\}.
\]

This condition follows, for example, from uniform \(\mathcal C^2\) regularity of the
relevant Brenier maps and Kantorovich potentials.
 {More precisely, assume that there exist}
\(\eta>0\), a compact convex set \(K\subset\mathbb R^d\), and constants
\(0<c<C<\infty\) such that \(Y\), \(h(\bm x)\), and all
\(\omega\in B_{\mathcal W_2}(h(\bm x),\eta)\) are absolutely continuous distributions
supported on \(K\), with densities bounded as
\(
    c\le \rho \le C\) on $K$.
{If} \(T_{\mu\to\nu}=\nabla\varphi_{\mu,\nu}\) denotes the Brenier map from
\(\mu\) to \(\nu\), then the corresponding potentials \(\varphi_{\mu,\nu}\)
are twice continuously differentiable on \(K\), and there exists a constant
\(C'<\infty\) such that
\(
    \sup_{\mu,\nu}
    \left(
        \|\nabla\varphi_{\mu,\nu}\|_{\infty}
        +
        \|\nabla^2\varphi_{\mu,\nu}\|_{\infty}
    \right)
    \le C',
\)
where the supremum is taken over all distributions \(\mu,\nu\) in the local
Wasserstein neighborhood under consideration.

Let
$ F_{\bm x}(t)=H(\omega_t,\bm x)$. Taylor's formula with integral remainder then gives
\[
    H(\omega,\bm x)-H\{h(\bm x),\bm x\}
    =
    F_{\bm x}(1)-F_{\bm x}(0)
    =
    F_{\bm x}'(0)+\int_0^1(1-t)F_{\bm x}''(t)\,dt .
\]
Since \(h(\bm x)\) is the local minimizer of \(H(\cdot,\bm x)\), the first-order
term vanishes:
\(    F_{\bm x}'(0)=0.
\)
{Moreover},
\[
    |F_{\bm x}''(t)|
    \le
    C_0
    \int_K \|T(u)-u\|^2\,dh(\bm x)(u)
    =
    C_0 \mathcal W_2^2\{\omega,h(\bm x)\}.
\]
Therefore,
\[
\begin{aligned}
    H(\omega,\bm x)-H\{h(\bm x),\bm x\}
    &\le
    \int_0^1(1-t)C_0\mathcal W_2^2\{\omega,h(\bm x)\}\,dt   =
    \frac{C_0}{2}\mathcal W_2^2\{\omega,h(\bm x)\}.
\end{aligned}
\]
%The lower bound follows from the local minimality of \(h(\bm x)\).
Hence, Assumption (A1) holds for multidimensional
Wasserstein responses under the above regular optimal-transport conditions.

\bibliographystyle{apalike}
\bibliography{ref}
\end{document}